\documentclass[aps,prx,twocolumn,superscriptaddress,longbibliography,nofootinbib]{revtex4-2}
\usepackage{amsfonts}
\usepackage{amsmath}
\usepackage{amssymb}
\usepackage{amsthm}
\usepackage{mathtools}
\usepackage{graphicx}
\usepackage{bm}
\usepackage{color}
\usepackage{comment}
\usepackage{mathrsfs}
\usepackage[]{mdframed}
\usepackage[colorlinks,bookmarks=true,citecolor=blue,linkcolor=red,urlcolor=blue]{hyperref}
\usepackage{appendix}
\usepackage{float}
\usepackage{booktabs}
\usepackage[export]{adjustbox}
\usepackage{xtab,afterpage}
\usepackage{algorithm}
\usepackage{algpseudocode}
\usepackage{setspace}
\usepackage{enumitem}
\usepackage{lipsum}
\usepackage[normalem]{ulem}

\allowdisplaybreaks

\newcommand{\RNum}[1]{\uppercase\expandafter{\romannumeral #1\relax}}

\theoremstyle{rmk}

\begin{document}
\title{
\textsc{ScarFinder}: a detector of optimal scar trajectories in quantum many-body dynamics
}
\author{Jie Ren}
\affiliation{School of Physics and Astronomy, University of Leeds, Leeds LS2 9JT, United Kingdom}

\author{Andrew Hallam}
\affiliation{School of Physics and Astronomy, University of Leeds, Leeds LS2 9JT, United Kingdom}

\author{Lei Ying}
\affiliation{School of Physics and Zhejiang Key Laboratory of Micro-nano Quantum Chips and Quantum Control, Zhejiang University, Hangzhou $310027$, China}

\author{Zlatko Papi\'c}
\affiliation{School of Physics and Astronomy, University of Leeds, Leeds LS2 9JT, United Kingdom}

\graphicspath{{./figs/}}

\begin{abstract}
Mechanisms that give rise to coherent quantum dynamics, such as quantum many-body scars, 
have recently attracted much interest as a way of controlling quantum chaos. However, identifying the presence of quantum scars in general many-body Hamiltonians remains an outstanding challenge. Here we introduce \textsc{ScarFinder}, a variational framework that reveals possible scar-like dynamics without prior knowledge of scar states or their algebraic structure, assuming only that such dynamics remain low in entanglement. By iteratively evolving and projecting states within a variational manifold, 
\textsc{ScarFinder} isolates scarred trajectories by suppressing thermal contributions. We validate the method on the analytically tractable spin-1 XY model, recovering the known scar dynamics, as well as the mixed field Ising model, where we capture and generalize the initial conditions previously associated with ``weak thermalization''. We then apply the method to the PXP model of Rydberg atom arrays, finding a previously unknown trajectory with  nearly-perfect revival dynamics in the thermodynamic limit. We also demonstrate that \textsc{ScarFinder} can efficiently identify the centers of stable islands in Poincar\'e sections of the mixed phase space that results from the projection of many-body quantum dynamics to a variational manifold. Our results establish \textsc{ScarFinder} as a powerful, model-agnostic tool for identifying and optimizing coherent dynamics in quantum many-body systems.
\end{abstract}

\maketitle

\section{Introduction}

Recent progress in quantum simulations~\cite{Bloch2012,Georgescu2014,Kjaergaard2020,Browaeys2020,MonroeRMP} has enabled \emph{in situ} monitoring of real-time dynamics and thermalization in isolated many-body quantum systems, allowing to directly probe foundational questions of quantum statistical mechanics. A powerful framework for describing thermalization in closed quantum systems is the eigenstate thermalization hypothesis (ETH)~\cite{DeutschETH, SrednickiETH, RigolNature, dAlessio2016,Ueda2020}. The ETH posits that in ``generic'' non-integrable systems -- those comprising many interacting degrees of freedom -- all eigenstates apart from spectral edges appear ``thermal'' in the sense of local observable expectation values. 
This elucidates how local observables equilibrate thermally as the entire system evolves unitarily.

However, many non-integrable systems once thought to fully obey the ETH are now understood to weakly violate it by hosting atypical eigenstates known as quantum many-body scars (QMBSs)~\cite{serbyn2021quantum,Moudgalya_2022,ChadranReview}. Some examples of systems that weakly violate the ETH include Rydberg atom arrays~\cite{bernien2017probing,turner2018weak}, the Heisenberg-type spin models~\cite{Moudgalya2018_2,Schecter2019,yang2024phantom}, ultracold atoms~\cite{Hudomal2020bosons,Zhao2020Optical,Desaules2021Tilted,scherg2021observing,Su2023,Adler2024,Evrard2024}, superconducting circuits~\cite{zhang2023many, Guo2023Origin, larsen2024experimentalprotocolobservingsingle}, and numerous other~\cite{ShiraishiMori,NeupertScars, Shibata202Onsager, McClarty2020, Lee2020Colored,  Mohapatra2023, kolb2023stability, Srivatsa2023, Desaules2023Schwinger, Halimeh2023robustquantummany, Gotta2023, Dooley2024Dual,pizzi2024quantumscarsmanybodysystems}. The ability to defy thermalization is of interest in quantum information processing~\cite{Dooley2021,Desaules2022QFI,Dooley2023}, e.g., QMBSs have been used to generate Greenberger-Horne-Zeilinger (GHZ) states~\cite{omran2019GHZ} and disorder-tunable entanglement far from equilibrium~\cite{Dong2023}.

A common trait of QMBS systems are long-lived, nonthermal dynamics when the system is initialized in special states with high overlap on QMBS eigenstates. In this work, we use the term ``scar eigenstate'' to refer to nonthermal eigenstates embedded in an otherwise thermal spectrum, and ``scar initial state'' to refer to an initial state that exhibits slow thermalization due to its large overlap with such eigenstates. Manifestations of slow thermalization include persistent oscillations in quench dynamics and slow entanglement growth~\cite{papić2021weakergodicitybreakinglens}, both of which are unexpected in fully chaotic systems. In some models, these features have an elegant theoretical description in terms of a so-called restricted su(2) spectrum generating algebra (RSGA)~\cite{moudgalya2020eta,Mark2020,ODea2020Tunnels,Bull2020}. For example, the RSGA approximately describes the QMBSs in the PXP model~\cite{FendleySachdev,Lesanovsky2012,turner2018weak,Ho2019,Choi2019,Surace2020, Khemani2019,lin2019exact, mondragon2021fate, Omiya2022,ivanov2025exactarealawscareigenstates}, an effective model of Rydberg atom arrays  in which persistent revivals have been observed in several experiments~\cite{bernien2017probing,bluvstein2021controlling,Su2023,zhao2024observationquantumthermalizationrestricted}. On the other hand, the theoretical landscape of QMBS models is much richer due to a multitude of exact constructions of non-thermalizing states and dynamics within chaotic many-body Hamiltonians~\cite{ShiraishiMori, pakrouski2020many, Mark2020eta, ren2021quasisymmetry, ren2022deformed, Buca2023, Moudgalya2024}. 

Despite much progress in the understanding of weak ETH violations, one key practical challenge remains: for a general model described by some unfamiliar Hamiltonian, how does one verify the existence of QMBSs or other kinds of non-thermalizing states? Such a task is non-trivial as QMBSs are typically associated with chaotic systems that lack integrability or conserved quantities, making them hard to treat analytically. In models with special structure, such as the PXP model, the QMBS eigenstates stand out as outliers with the lowest entanglement entropy in the spectrum~\cite{Turner2018PRB} and the initial states associated with them can be inferred based on semiclassical intuition~\cite{Ho2019,Michailidis2020,Daniel2023}. However, this does not easily generalize to other models, including the PXP-type models with longer-range constraints~\cite{kerschbaumer2024quantummanybodyscarspxp}. Similarly, numerical tools such as exact diagonalization or tensor network methods, are limited by system size or entanglement, which restricts their applicability in general. 

\begin{figure}
    \centering
    \includegraphics[width=1\linewidth]{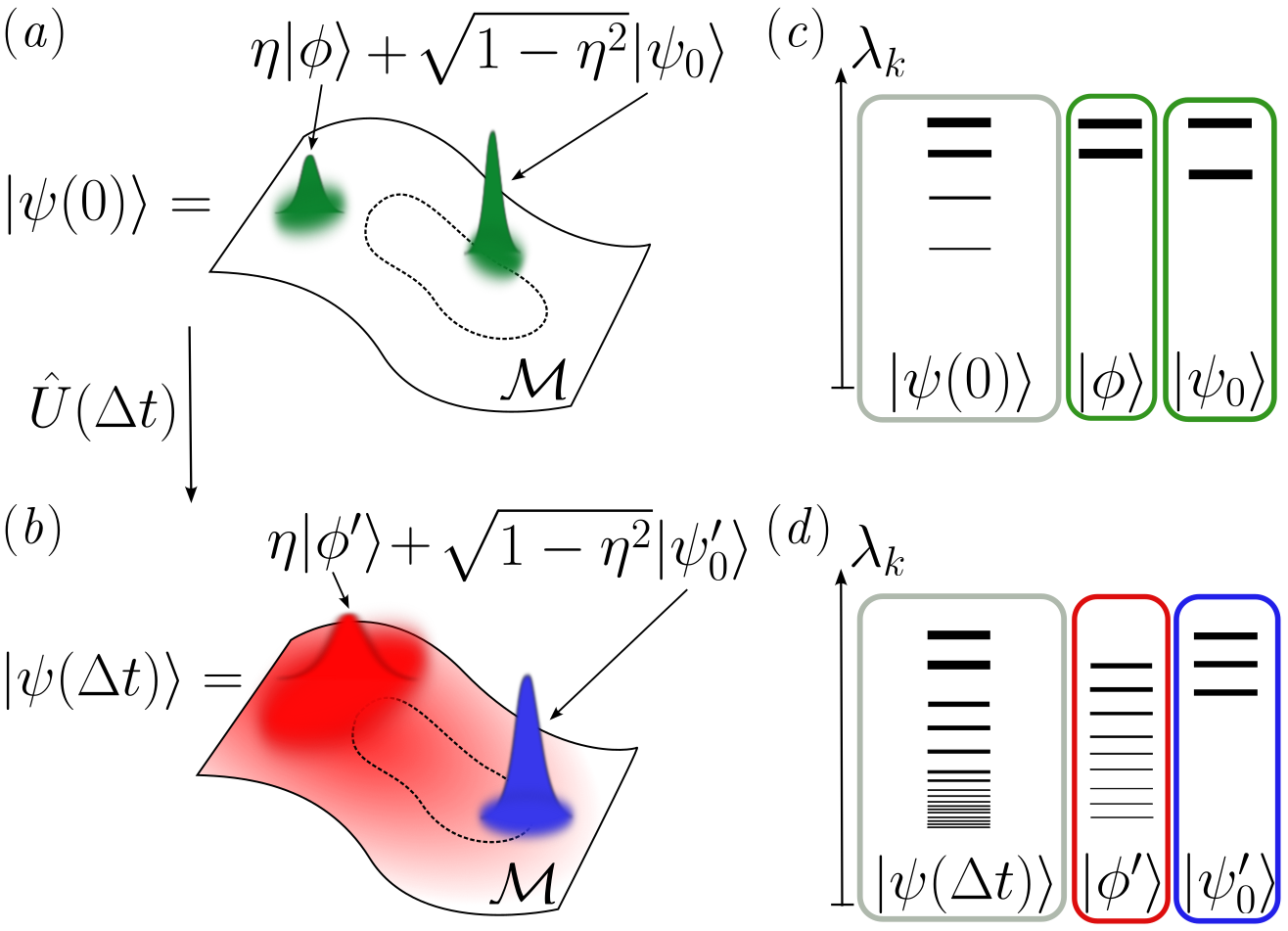}
    \caption{Schematic illustration of a single step of the \textsc{ScarFinder} algorithm. (a) The initial state \( |\psi(0)\rangle \), lying on a low-entanglement variational manifold $\mathcal{M}$, is assumed to be a superposition of a scar component \( |\psi_0\rangle \), which evolves periodically (dashed line), and a generic component \( |\phi\rangle \). (b) After a unitary time evolution step \( \hat U(\Delta t) \), the state becomes \( |\psi(\Delta t)\rangle \), with the superposition preserved. However, while the scar component \( |\psi_0'\rangle \) remains low-entangled, the generic component \( |\phi'\rangle \) becomes highly entangled and delocalized. (c) At \( t = 0 \), both components exhibit low entanglement, characterized by a small number of dominant Schmidt coefficients $\lambda_k$. (d) The delocalization of $|\phi^\prime\rangle$ leads to a broader distribution of Schmidt coefficients. This entanglement difference enables the separation of the two components in \(|\psi(\Delta t)\rangle\): the leading Schmidt coefficients are dominated by the scar part, while the subleading terms largely originate from the thermal component. The projection step in \textsc{ScarFinder} leverages this structure to suppress thermal contributions and reinforce the scar trajectory.}
    \label{fig:sketch}
\end{figure}

In this work, we introduce the \textsc{ScarFinder} method which uncovers scar-like dynamics in general chaotic systems. The key principle behind \textsc{ScarFinder} is that scarred dynamics are typically confined to a low-entanglement submanifold  of the Hilbert space. This allows to systematically search for scar trajectories by iteratively evolving and projecting states within a suitable variational manifold, such as matrix product states (MPS) with fixed bond dimension.  The only required inputs for \textsc{ScarFinder} are the system's Hamiltonian and a choice of this variational manifold that harbors the low-entanglement dynamics. Crucially, this approach requires no prior knowledge of scar eigenstates or their algebraic structure, and depends only on the existence of a suitable low entanglement variational manifold. The algorithm then leverages the dynamical separation between scarred and thermal components that emerges during time evolution within this manifold: while thermal components spread rapidly and exit the manifold, the scar components remain coherent and confined to the manifold, see Fig.~\ref{fig:sketch}. By combining short-time evolution with projection back to the manifold, \textsc{ScarFinder} effectively suppresses thermal contributions and converges toward periodic trajectories.

The remainder of this paper is organized as follows. In Sec.~\ref{sec:algorithm} we motivate the \textsc{ScarFinder} method and outline the structure of the algorithm. In Sec.~\ref{sec:spin-1-xy}, to facilitate understanding, we validate the algorithm using an analytically-tractable example of the spin-1 XY model that hosts exact QMBS states~\cite{Schecter2019}. We demonstrate that our method recovers the known scar trajectories without any input of their algebraic construction, and we discuss in detail the convergence of the method. In Sec.~\ref{sec:PXP}, we apply the method to the PXP model, where we identify a new scar trajectory that enhances the previously known trajectory associated with the N\'eel initial state and results in robust -- almost perfect -- revivals in the thermodynamic limit. This highlights the power of the method both in discovering scars as well as optimizing known scar dynamics in systems without exact solutions. 
Finally, in Sec.~\ref{sec:poincare} we show that 
\textsc{ScarFinder} can reveal regions of regular dynamics within ``mixed phase space'' of many-body quantum systems by projecting their evolution onto a low-dimensional variational manifold, where classical structures like Kolmogorov-Arnold-Moser (KAM) tori and Poincar\'e sections can be directly identified with high efficiency. Our conclusions are presented in Sec.~\ref{sec:conc}, while Appendices contain further technical details of the algorithm, its pedagogical demonstration, and an application beyond QMBS systems, where we successfully reproduce and generalize the findings of Refs.~\cite{Banuls2011,LinMotrunich2017} for the mixed field Ising model. 

\section{The \textsc{ScarFinder} algorithm}\label{sec:algorithm}

Identifying initial states that lead to periodic or otherwise non-thermal dynamics within chaotic many-body systems is an important goal in understanding the applicability of the ETH and the mechanisms of its breakdown. Beyond QMBSs, recent examples of time crystals~\cite{PhysRevLett.109.160401,khemani2019briefhistorytimecrystals,RevModPhys.95.031001} and the quantum Mpemba effect~\cite{ares2025quantummpembaeffects,bhore2025quantummpembaeffectglobal} further highlight the importance of this task. In the former case, states exhibit periodic motion by spontaneously breaking time-translation symmetry, while in the latter the system's relaxation dynamics depend counterintuitively on the effective temperature of the initial state. Despite their distinct origins, these phenomena share a common difficulty: how to pinpoint the special initial conditions in an exponentially large state space. Motivated by this
need for a systematic method, below we introduce the \textsc{ScarFinder} algorithm and present an intuitive illustration of its underlying principle.

\subsection{The algorithm and its motivation}\label{sec:alg_motivation}

The core idea of \textsc{ScarFinder} is illustrated schematically in Fig.~\ref{fig:sketch}. Suppose that for a given many-body Hamiltonian $\hat H$, there exists a variational manifold \( \mathcal{M} \) that captures the scar dynamics (either exactly or to a good approximation). Specifically, let us assume there exists a state \( |\psi_0\rangle \in \mathcal{M} \) whose time evolution remains confined to the manifold and traces out a periodic trajectory \( |\psi_0(\theta)\rangle \), parametrized by a variable $\theta$. 

Consider now a state slightly perturbed away from the ideal scar trajectory. As illustrated in Fig.~\ref{fig:sketch}(a), the state can be written as a superposition of a generic (non-scar) component $|\phi\rangle$ and a scar component $|\psi_0(\theta)\rangle$:
\begin{equation}\label{eq:scar_decomp}
    |\psi\rangle = \eta |\phi\rangle+\sqrt{1 - \eta^2} |\psi_0(\theta)\rangle,
\end{equation}
where both components belong to \( \mathcal{M} \) and $\eta$ is the weight of the non-scar component. To gain some intuition, let us first assume that $\eta$ is small, such that the initial state is in the vicinity of the periodic orbit. Since both components lie within the manifold, separating them typically requires detailed knowledge of \( |\psi_0\rangle \). However, the key insight of \textsc{ScarFinder} is that the distinct dynamical behavior of these components enables their separation without such knowledge.

After evolving the state under the Hamiltonian for a sufficiently long time \( \Delta t \) under $\hat U(\Delta t) =\exp(-i\hat{H} \Delta t)$, we obtain:
\begin{equation}\label{eq:timeevolution}
    |\psi(\Delta t)\rangle  %\hat U(\Delta t) |\psi\rangle 
    = \eta\, \hat U(\Delta t)|\phi \rangle + \sqrt{1 - \eta^2} \, \hat U(\Delta t) |\psi_0\rangle,
\end{equation}
In scarred systems, it is natural to expect that the second, component in the above equation remains coherent and low-entangled, while the generic component, $\hat U(\Delta t)|\phi\rangle$, typically thermalizes and spreads across the Hilbert space.
Thus, after projection \( \hat{P}_\mathcal{M} \) back onto the manifold \( \mathcal{M} \), we obtain:
\begin{equation}\label{eq:projectionM}
    \hat{P}_\mathcal{M}|\psi(\Delta t)\rangle = %\hat{P}_\mathcal{M} e^{-i\hat{H} \Delta t} |\psi\rangle = 
    \eta' \, |\phi'\rangle + \sqrt{1 - (\eta^\prime)^2} \,|\psi_0(\theta')\rangle,
\end{equation}
where the prime denotes components projected to $\mathcal{M}$ and  \( \eta' \leq \eta \) indicates the expected reduced weight of the thermal component, see Fig.~\ref{fig:sketch}(b). While we do not have a proof of Eq.~(\ref{eq:projectionM}) -- and indeed do not expect it to be generally satisfied for an arbitrary $\hat H$ -- in Sec.~\ref{sec:spin-1-xy} and Appendix~\ref{apx:finite_xy} we will demonstrate its validity in physical QMBS models. 

The difference between the two components in Eq.~(\ref{eq:projectionM}) can be revealed by performing the Schmidt decomposition across a bipartition $A{\cup}B$ in the middle of the system:
\begin{equation}\label{eq:schmidt_decomp}
    |\psi\rangle = \sum_{k} \lambda_k |\psi^A_k\rangle |\psi^B_k\rangle,
\end{equation}
where \( \{\lambda_k\} \) are the Schmidt coefficients. Applying Eq.~(\ref{eq:schmidt_decomp}) to the scar component \( |\psi_0(\theta')\rangle \) gives a sharply concentrated set of Schmidt values with only a few dominant ones, while the thermalizing component \( |\phi'\rangle \) yields a broad distribution of smaller Schmidt values, as illustrated in Fig.~\ref{fig:sketch}(c)-(d).
As a result, even in the combined state \( |\psi(\Delta t)\rangle \), the Schmidt spectrum exhibits a `two-component' structure: the dominant coefficients are governed by the scar component, while the tail contains contributions from the thermal component. 

In summary, we conjecture that each iteration of the evolution-projection cycle, implemented via repeated applications of Eqs.~(\ref{eq:timeevolution})-(\ref{eq:projectionM}), edges the state closer to the scar trajectory. 
The natural separation of the Schmidt coefficients motivates a truncation procedure  that effectively suppresses the non-scar part, with the structure of the full 
\textsc{ScarFinder} presented in Algorithm~\ref{alg:findscar}.   
In the following Sec.~\ref{sec:alg_imp}, we discuss the different steps and their implementation in detail. 

\subsection{Implementation}\label{sec:alg_imp}

Here, we highlight some important practical aspects of \textsc{ScarFinder}. First, the manifold \( \mathcal{M} \) can be arbitrary. Throughout this work, we will focus on spin chains and cases where \( \mathcal{M} \) is spanned by infinite matrix product states (iMPS)~\cite{CiracRMP} with an \( n \)-site unit cell:
\begin{equation}
    \big|\psi ([A_1,\!\cdots\!,A_n])\big\rangle  \\
    = \sum_{\bm s} \Big(\cdots\! A^{[s_i]}_1 A^{[s_{i+1}]}_2 \!\cdots\! A^{[s_{i+n-1}]}_n \!\cdots\! \Big)|\bm s\rangle,
\end{equation}
where $A_i$ are matrices of maximum bond dimension $\chi$ and $|\bm s\rangle$ denotes basis states. 
This manifold faithfully captures low-entanglement states in the thermodynamic limit. Provided that the target dynamics is captured by such states, we shall not require any further knowledge of the scar structure.

\begin{algorithm}[H]
\caption{\textsc{ScarFinder} algorithm}
\label{alg:findscar}
{\bf Input:}  An appropriate variational manifold $\mathcal{M}$, Hamiltonian $\hat{H}$, projection time step $\Delta t$, number of iteration steps $N_\text{step}$, target energy density $E_\text{target}$; \\
{\bf Output:} A potential scarred state $|\psi\rangle \in \mathcal{M}$;
{\setstretch{1.35}
\begin{algorithmic}[1]
\State {\bf Initialize:} A trial state $|\psi\rangle \in \mathcal{M}$; 
\For{$n=1:N_\text{step}$}
\State\label{alg:state_evo} Evolve the $|\psi\rangle$ to $|\psi(\Delta t)\rangle \coloneq \exp(-i\hat{H}\Delta t)|\psi\rangle$; 
\State\label{alg:projection} Project back onto $\mathcal{M}$, $|\psi^\prime\rangle \coloneq \hat P_\mathcal{M}|\psi(\Delta t)\rangle$;
\State \hangindent2.5em\hangafter1Enforce energy conservation if the energy of $|\psi^\prime\rangle$ deviates from $E_\text{target}$, and any crucial symmetries or constraints, resulting in $|\psi^{\prime\prime}\rangle$;
\State Update $|\psi\rangle \coloneq |\psi^{\prime\prime}\rangle$;
\EndFor
\State \Return $|\psi\rangle$
\end{algorithmic}
}
\end{algorithm}

Previously, we motivated the algorithm heuristically, assuming that the initial state is close to the scar state ($\eta \ll 1$) and the projection time step \( \Delta t \) is large. In practice, neither assumption necessarily holds: we do not know the scar state \emph{a priori}, while large time steps are computationally costly. Regarding $\eta$, if we do not assume prior knowledge about the structure of the periodic orbit, the natural strategy is sampling over randomly chosen states within $\mathcal{M}$. 
Although convergence is not guaranteed for every initial condition, our algorithm exhibits a high success rate for all models we tested. It is important to note that even though the scar states are known in these models, we independently obtain them using the \textsc{ScarFinder} algorithm. We found that only a few random initializations are needed to reliably recover these scar trajectories, and we have not encountered cases that require extensive sampling, indicating the algorithm's low sampling complexity in practice.

The time-evolution step (line~\ref{alg:state_evo}) of the algorithm can be conveniently performed using standard methods, e.g., both infinite time-evolving block decimation (iTEBD)~\cite{VidalTEBD} and time-dependent variational principle (TDVP)~\cite{haegeman2011time,Haegeman2014geometry} can be used for MPS manifolds. 
These methods favor smaller $\Delta t$ due to the entanglement buildup, which increases the complexity of the simulation. On the other hand, \textsc{ScarFinder} naturally prefers large $\Delta t$, which allows the thermal component $|\phi^\prime\rangle$ to sufficiently delocalize over $\mathcal{M}$. 
Note that we will distinguish between \(\Delta t\) and \(t\): \(\Delta t\) refers to the projection interval used during the \textsc{ScarFinder} optimization, while \(t\) denotes the real-time evolution starting from the optimized initial state. 
In general, larger values of $\Delta t$ enhance the separation between thermal and scar components during time evolution. Thus, one would choose $\Delta t$ to be as large as possible, given the available computational resources. In several examples considered below, we find that the entanglement-based separation persists qualitatively for moderate values of \( \Delta t \) that are commonly used in MPS-based methods for time evolution.

The projection step (line~\ref{alg:projection}) 
is the most subtle; if this is not performed carefully, the algorithm may converge to a trivial low-entanglement state, such as the ground state, rather than the desired scar trajectory at a non-zero energy density in the many-body spectrum. The projection may inadvertently break important conservation laws, e.g., for an iMPS manifold, bond truncation can break energy conservation. 
To address this, we perform imaginary-time evolution to systematically reduce errors arising from bond truncation after each projection step. For a specific target energy \( E_{\text{target}} \), we monitor the energy deviation, 
$\Delta E = \langle \hat{H} \rangle - E_{\text{target}}$, after each truncation step.
We then define a small imaginary-time step \( d\tau = \Delta E/n \) (typically choosing \( n = 10 \)) and evolve the state within the fixed-bond-dimension MPS manifold:
\begin{equation}\label{eq:energyfix}
    |\psi_0\rangle = |\psi\rangle, \quad 
    |\psi_n\rangle = \hat{P}_\mathcal{M}\left[e^{-\hat H d\tau } |\psi_{n-1}\rangle\right].
\end{equation}
This procedure yields a sequence of energies 
$E_i=\langle\psi_i|\hat H|\psi_i\rangle$ with $i=1,2,\dots,n$. We 
track this sequence until an energy value closest to the target \( E_{\text{target}} \) is found. However, since this optimization is constrained to a manifold with fixed bond dimension, achieving exact convergence to the desired energy may not always be feasible. If we observe that the energy difference consistently increases or fails to approach zero after multiple steps, the optimization attempt is halted and considered unsuccessful for that initial configuration.

Finally, after converging the results for one value of $\chi$, one might want to test their sensitivity to increasing $\chi$. The convergence of the results  with $\chi$  is not guaranteed to be monotonic due to optimization instabilities or getting trapped in local minima. Nevertheless, in all scar models we studied in this paper, which do not admit an exact scar state within a fixed-bond-dimension MPS manifold, we found that increasing the bond dimension $\chi$ generally improved the performance of the algorithm. In the following Sec.~\ref{sec:spin-1-xy}, we will apply the method to the analytically-tractable example of the spin-1 XY model~\cite{Schecter2019}, which will allow to concretely illustrate the above steps and build intuition about the inner workings of the algorithm.

\section{Spin-1 XY Model}\label{sec:spin-1-xy}

One of the simplest and most studied models that contain analytically exact QMBSs is the spin-1 XY chain~\cite{Schecter2019,Chattopadhyay2020}:
\begin{equation}\label{eq:spin-1XY}
\begin{aligned}
    \hat{H} &= \sum_j \left( \hat{S}_j^x \hat{S}_{j+1}^x 
	+ \hat{S}_j^y \hat{S}_{j+1}^y \right) - h \sum_j \hat{S}^z_j + \hat{V},
\end{aligned}
\end{equation}
where $\hat S_j^\alpha$ are the standard spin-1 operators with $\alpha=x,y,z$ on site $j$, $h$ is the external magnetic field in the $z$-direction, and $\hat V$ is a perturbation term that breaks integrability and any symmetries that are not essential for QMBS states. 
In previous works~\cite{Schecter2019,Chattopadhyay2020}, it was shown that the model (\ref{eq:spin-1XY}) hosts two families of QMBS eigenstates, known as the Type-1 and Type-2 scar towers. The two types are distinguished by the initial state they overlap with; as we detail below, the Type-I tower overlaps with a product state, while the other, Type-II, tower has overlap with an MPS state of bond dimension $\chi=2$. From the perspective of \textsc{ScarFinder}, the algorithm performs similarly in the two cases, hence in the remainder of this section we focus on Type-1 scar tower, while the results for Type-2 are relegated to Appendix~\ref{apx:xy2}.  

To simplify the analysis, we will choose $\hat V$ that breaks both integrability and the U(1) symmetry of the model, i.e., the conservation of magnetization $\sum_j S_j^z$, while preserving the desired scar structure. For the Type-1 scar tower, this is achieved by the following perturbation: 
\begin{equation}\label{eq:V1}
    \hat V_1 = \sum_j \hat P^{0}_j \hat S^x_{j+1}, \;\;\; \hat P_j^{0} = \hat{\mathbb{I}} - (\hat S_j^z)^2= |0_j\rangle\langle0_j|.
\end{equation}
For the perturbation $\hat V_1$ in Eq.~(\ref{eq:V1}), the spin-1 XY model hosts an exact tower of QMBS eigenstates:
\begin{equation}\label{eq:Type1QMBS}
    |\psi_n\rangle = (\hat{Q}^+)^n |-\cdots -\rangle,
\end{equation}
generated by the ladder operator \(\hat{Q}^+\):
\begin{equation}
    \hat{Q}^+ = \sum_j (-1)^j (\hat{S}_j^+)^2.
\end{equation}
Here, $|+\rangle$ and $|-\rangle$ denote the highest and lowest weight spin-1 basis states, respectively.

The scar eigenstates $|\psi_n\rangle$ in Eq.~(\ref{eq:Type1QMBS}) can be coherently combined to form special initial states that exhibit persistent, nonthermal dynamics under unitary evolution~\cite{Schecter2019}: 
\begin{equation}\label{eq:xy1_scar_state}
    |\psi_0(\theta,\xi)\rangle = e^{-i\theta\sum_j\hat{S}_j^z}e^{\xi \hat Q^+}|-\cdots -\rangle.
\end{equation}
To explicitly illustrate the working principle of the \textsc{ScarFinder} algorithm, we consider an imperfect initial state parameterized by a real number $\alpha$:
\begin{equation}\label{eq:imperfect_scar}
    |\psi_\alpha\rangle = \exp\Big[-i\alpha\sum_j(-1)^j\hat S^z_j\Big]\big|\psi_0(0,1)\big\rangle.
\end{equation}
When $\alpha = 0$, the initial state is identical to the scar state $|\psi_0(0,1)\rangle$. Thus, nonzero values of $\alpha$ serve as a measure of the departure from the ideal scar trajectory.

A subtlety arises in infinite systems: the intuitive decomposition introduced earlier in Eq.~(\ref{eq:scar_decomp}) is ill-defined for iMPS due to the challenge of defining a proper norm. To explicitly illustrate convergence in terms of the parameter $\eta$, we provide a pedagogical example in Appendix~\ref{apx:finite_xy}, which utilizes the exact scar structure in a finite size $L=6$. For infinite systems, instead of the global norm, we quantify the overlap between two iMPS wavefunctions, $\psi(A_1)$ and $\psi(A_2)$, using the largest eigenvalue of the generalized transfer matrix: 
\begin{equation}
    \mathcal N[\psi(A_1),\psi(A_2)] = \max \big|\operatorname{eigvals}[T(A_1,A_2)]\big|.
\end{equation}
We refer to this quantity as the logarithmic fidelity. Note that the fidelity revival implies recurrence of all local observables. Since the scar states form a continuous submanifold \(\mathcal{S}\), we introduce a fidelity measure -- the ``scar fidelity'' -- to quantify how closely a given state \(\psi\) aligns with \(\mathcal{S}\):
\begin{equation}\label{eq:scarfid}
    F_\mathcal{S}[\psi] = \max_{s \in \mathcal{S}} \mathcal{N}[s,\psi].
\end{equation}
We emphasize that this fidelity serves purely as a diagnostic tool for evaluating convergence toward the scar manifold. The \textsc{ScarFinder} algorithm itself requires no explicit information about the scar subspace, as it operates solely within a generic manifold of fixed-bond-dimension MPS.

\begin{figure}[tb]
    \centering
    \includegraphics[width=\linewidth]{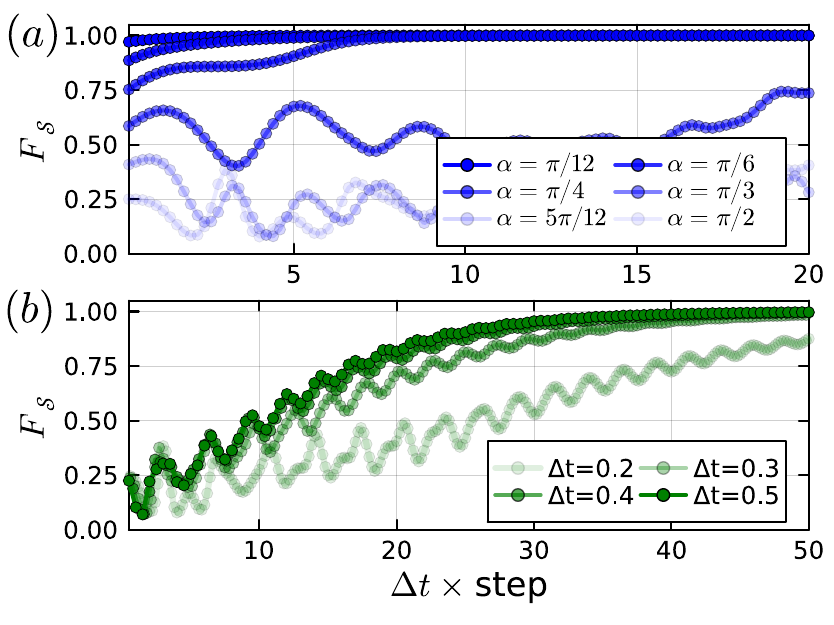}
    \caption{
        Convergence of scar fidelity \(F_\mathcal{S}\) in Eq.~(\ref{eq:scarfid}) for initial states \( |\psi_\alpha\rangle \) in Eq.~(\ref{eq:imperfect_scar}), simulated using iTEBD. 
        (a) For a fixed projection time step \(\Delta t=0.2\), states quickly converge to the scar trajectory for smaller values of \(\alpha\), whereas larger \(\alpha\) slows convergence.
        (b) For fixed deviation \(\alpha=\pi/2\), larger values of \(\Delta t\) generally accelerate convergence. 
    }
    \label{fig:xy1_fs}
\end{figure}

In Fig.~\ref{fig:xy1_fs}, we fix the value of field strength at \(h=1.0\) and perform numerical simulations starting from \( |\psi_\alpha\rangle \), which we represent as iMPS with a two-site unit cell. For a fixed projection time step \(\Delta t=0.2\), we observe rapid convergence for smaller values of \(\alpha\), whereas larger deviations (\(\alpha\approx\pi/2\)) result in slower convergence, indicating an initial-state dependence. Conversely, as shown in Fig.~\ref{fig:xy1_fs}(b), by fixing \(\alpha=\pi/2\) and varying \(\Delta t\), we confirm that larger evolution intervals enhance convergence speed. Thus, we conclude that there is a tradeoff between $\alpha$ and $\Delta t$: for poor initial guesses that land far from the scar trajectory in $\mathcal{M}$, we need to evolve the system for a longer time $\Delta t$. 

We remark that during iterative applications of \textsc{ScarFinder}, the state's energy may drift due to bond truncation. Although the energy density of the scar state in Eq.~(\ref{eq:xy1_scar_state}) covers the interval \([-h,+h]\), uncontrolled energy drift could complicate precise trajectory selection. This is why it is essential to enforce energy targeting, as previously explained in Sec.~\ref{sec:alg_imp}. 

\begin{figure}[tb]
    \centering
    \includegraphics[width=\linewidth]{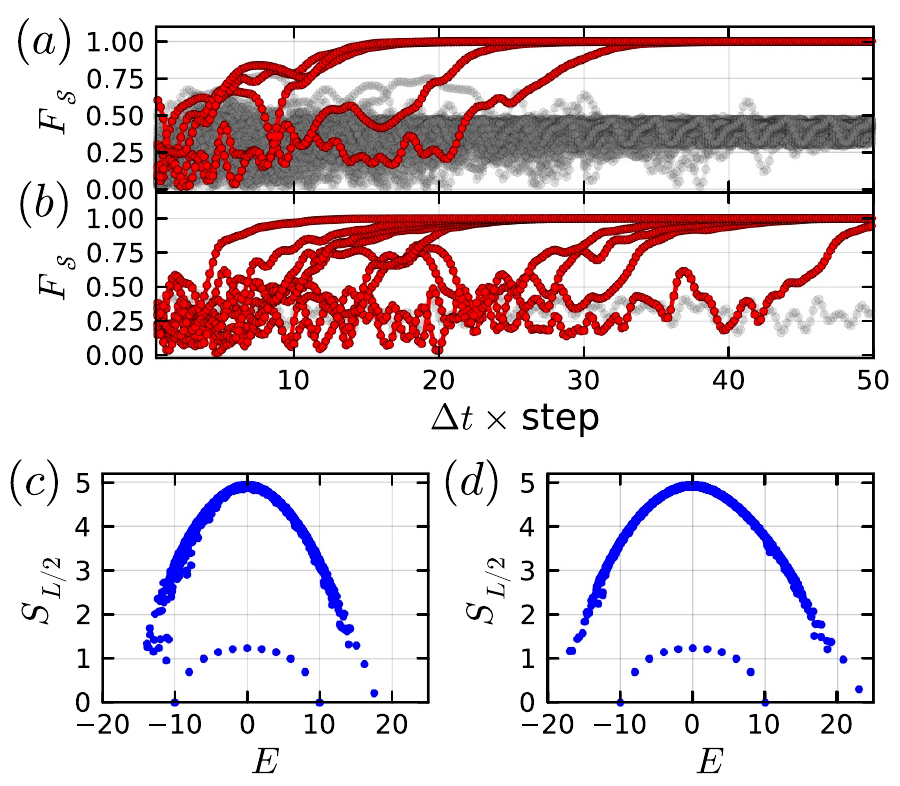}
    \caption{
    Convergence behavior of \textsc{ScarFinder} initialized from random states at target energy \(E_\text{target}=0\), with $\Delta t=0.2$. Energy conservation is explicitly enforced during the algorithm's iterations.
    (a) For the  \(\hat V=\hat V_1\) perturbation in Eq.~(\ref{eq:V1}), among 100 independent trials, only 5 successfully converge. 
    (b) For the  \(\hat V = \hat V_1'\) perturbation in Eq.~(\ref{eq:V1prime}), 9 out of 10 trials converge successfully.
    (c),(d): Entanglement entropy of eigenstates for (c) \(\hat V = \hat V_1\) and (d) \(\hat V = \hat V_1'\). Improved convergence in (b) correlates with a clearer separation between scar and thermal states in the entanglement entropy spectrum. Data in panels (a),(b) are obtained using iTEBD with a maximal bond dimension \(\chi=16\) and evolution step \(d t=0.01\). Panels (c)-(d) are exact diagonalization results for a finite system $L=10$, combining momentum sectors \(k=0\) and \(k=\pi\).
    }
    \label{fig:xy1_random}
\end{figure}

To illustrate sensitivity to initial conditions, we run the energy-corrected \textsc{ScarFinder} algorithm with randomly initialized states  in Fig.~\ref{fig:xy1_random}(a). For the perturbation \(\hat{V}_1\), the success rate is low, highlighting significant initial-state dependence. The reason for this is the simplicity of $\hat V_1$ perturbation, which leaves some residual structure in the non-scar eigenstates. To test the assumptions underlying the \textsc{ScarFinder} algorithm, we next consider a more general perturbation 
\begin{equation}\label{eq:V1prime}
    \hat{V}_1' = \sum_j \hat{P}^0_j \left[\sum_{i=1}^8 \sin(100\,i)\hat \lambda^{(i)}_{j+1}\right],
\end{equation}
where \(\hat\lambda^{(i)}\) are Gell-Mann matrices. This perturbation can be derived from the quantum inverse method, as detailed in Appendix~\ref{apx:parent-ham}.
While neither perturbation is fully generic due to this constraint, the second one includes a broader range of local terms and thus breaks more of the model’s residual structure. As shown in Fig.~\ref{fig:xy1_random}(b), the $\hat V_1'$ perturbation indeed substantially improves the convergence rate.

We attribute the difference in performance to residual symmetry-related structures in the eigenstate spectrum. 
To characterize the spectrum, we perform exact diagonalization of finite-size chains with periodic boundary condition, which are expected to approximately capture the main features of infinite systems. For each of the energy eigenstates, we perform the Schmidt decomposition, Eq.~(\ref{eq:schmidt_decomp}), and we evaluate the entanglement entropy
\begin{equation}\label{eq:SE}
    S_{A}=-\sum_k \lambda_k^2 \ln \lambda_k^2,
\end{equation}
where $A$ is assumed to be one half of the chain.
While the perturbation \(\hat V_1\) breaks the original U(1) symmetry, it does not entirely erase the associated spectral features, as evident from the exact diagonalization data in Fig.~\ref{fig:xy1_random}(c). Consequently, the thermalizing component $|\phi'\rangle$ does not fully exhibit the generic entanglement structure assumed in Sec.~\ref{sec:algorithm}, weakening the convergence performance. A more generic perturbation, such as \(\hat V_1'\), better suppresses the residual structures, resulting in a clearer separation between scar and thermal eigenstates [Fig.~\ref{fig:xy1_random}(d)] with significantly improved convergence.

Finally, for the spin-1 XY model, in addition to the Type-1 QMBS states discussed above, there exists a distinct Type-2 scar tower~\cite{Schecter2019,Chattopadhyay2020}. 
This Type-2 scar tower features a weakly-entangled scar trajectory described by MPS. In Appendix~\ref{apx:xy2} we demonstrate that the \textsc{ScarFinder} successfully obtains the Type-2 trajectory, with similar convergence criteria as in the Type-1 case illustrated above.

\section{PXP model}
\label{sec:PXP}

In this section we turn our attention to the experimentally relevant example of QMBSs in the PXP model, which is natively realized in Rydberg atom arrays~\cite{bernien2017probing,bluvstein2021controlling} and has also been engineered using tilt potential in a Bose-Hubbard optical lattice~\cite{Su2023}. The PXP model~\cite{FendleySachdev,Lesanovsky2012,turner2018weak} represents a one-dimensional spin-1/2 chain with the Hamiltonian:
\begin{equation}\label{Eq:PXP model}
    \hat H_\mathrm{PXP} = \Omega\sum_{j} \hat P_{j-1}\hat \sigma_{j}^x \hat P_{j+1},
\end{equation}
 where $\Omega=1$ is the Rabi frequency, $\hat \sigma_j^x$ is the standard Pauli-$x$ operator on site $j$, and $\hat P_j =(1-\hat \sigma_j^z)/2$ is the projector on the local $|{\downarrow}\rangle$ state. The PXP model describes constrained flipping of atoms in the Rydberg blockade regime~\cite{Browaeys2020}: each atom can flip only if both of its neighbors are in the $|\downarrow\rangle$ state. Hence, neighboring excitations, such as  $|{\cdots}{\uparrow}{\uparrow}{\cdots}\rangle$, are energetically forbidden, which imposes a constraint on the dynamics. 

The PXP model is known to be chaotic~\cite{turner2018weak}, while at the same time it also hosts a small number of non-thermal QMBS eigenstates which are evenly distributed in energy and possess anomalously low entanglement entropy~\cite{turner2018weak, Turner2018PRB, IadecolaSchecterXu2019, lin2019exact}. 
These QMBS eigenstates have high overlap with the  $|\mathbb{Z}_2\rangle \equiv |{\downarrow}{\uparrow}{\downarrow}{\uparrow}{\cdots}\rangle$ state, leading to a suppressed growth of entanglement entropy and periodic revivals in local observables when the system is quenched from that state. The resulting scar dynamics were shown to have an elegant semiclassical explanation in terms of a variational manifold $\mathcal{M}$ spanned by $\chi=2$ MPS states~\cite{Ho2019}. Equivalently, the scar dynamics can be interpreted as precession of a collective spin in an approximate su(2) RSGA picture~\cite{Turner2018PRB,Choi2019,Bull2020,Omiya2022}. The latter approach has also allowed to construct deformations of the PXP model which lead to nearly perfect revivals from the $|\mathbb{Z}_2\rangle$ initial state~\cite{Choi2019,Khemani2019,Bull2020,Omiya2022}. 

The PXP revival dynamics is understood to be a many-body analog of scar phenomena in stadium billiards, which are quantum remnants of classical, unstable periodic orbits~\cite{Heller84, HellerLesHouches, Bogomolny1988, Berry1989Scars}.  Nevertheless, some important questions remain open. For example, the QMBS subspaces beyond those associated with $|\mathbb{Z}_2\rangle$ are poorly understood, a simple example being the reviving $|\mathbb{Z}_3\rangle\equiv |{\downarrow}{\downarrow}{\uparrow}{\downarrow}{\downarrow}{\uparrow}{\cdots}\rangle$ state with a repeated 3-site unit cell~\cite{Turner2018PRB}. This leads to the question: can one do better than $|\mathbb{Z}_2\rangle$ and $|\mathbb{Z}_3\rangle$ states, i.e., can one construct low-entangled initial states that have more pronounced revivals than these product states? Specifically, for the undeformed PXP model, is it possible to find initial states that exhibit nearly perfect revival dynamics and almost no entanglement growth in the thermodynamic limit?  In the remainder of this section, we demonstrate that \textsc{ScarFinder} algorithm can provide useful insights into these questions.

\subsection{Two-site unit cell}

When directly applying the \textsc{ScarFinder} algorithm to the PXP Hamiltonian in Eq.~(\ref{Eq:PXP model}), a subtlety arises due to the Rydberg blockade constraint: the state \( \left|\uparrow\uparrow\right\rangle \) is forbidden on adjacent sites, therefore the expectation value of the projector \( \left|\uparrow\uparrow\right\rangle\left\langle\uparrow\uparrow\right| \) must vanish throughout the dynamics. However, bond truncation of iMPS during the simulation can violate this constraint. To mitigate this, we introduce an artificial penalty term:
\begin{equation}
    \hat H_\text{PXP}^\prime = \hat H_\text{PXP} - i\mu \sum_j \left|\uparrow\uparrow\right\rangle_{j,j+1}\left\langle\uparrow\uparrow\right|_{j,j+1},
\end{equation}
where \(\mu\) is a large positive constant (we set \(\mu = 100\) in our simulations). The imaginary coefficient then suppresses the unphysical components that could be generated during time evolution.

\begin{figure}[tb]
    \centering
    \includegraphics[width=\linewidth]{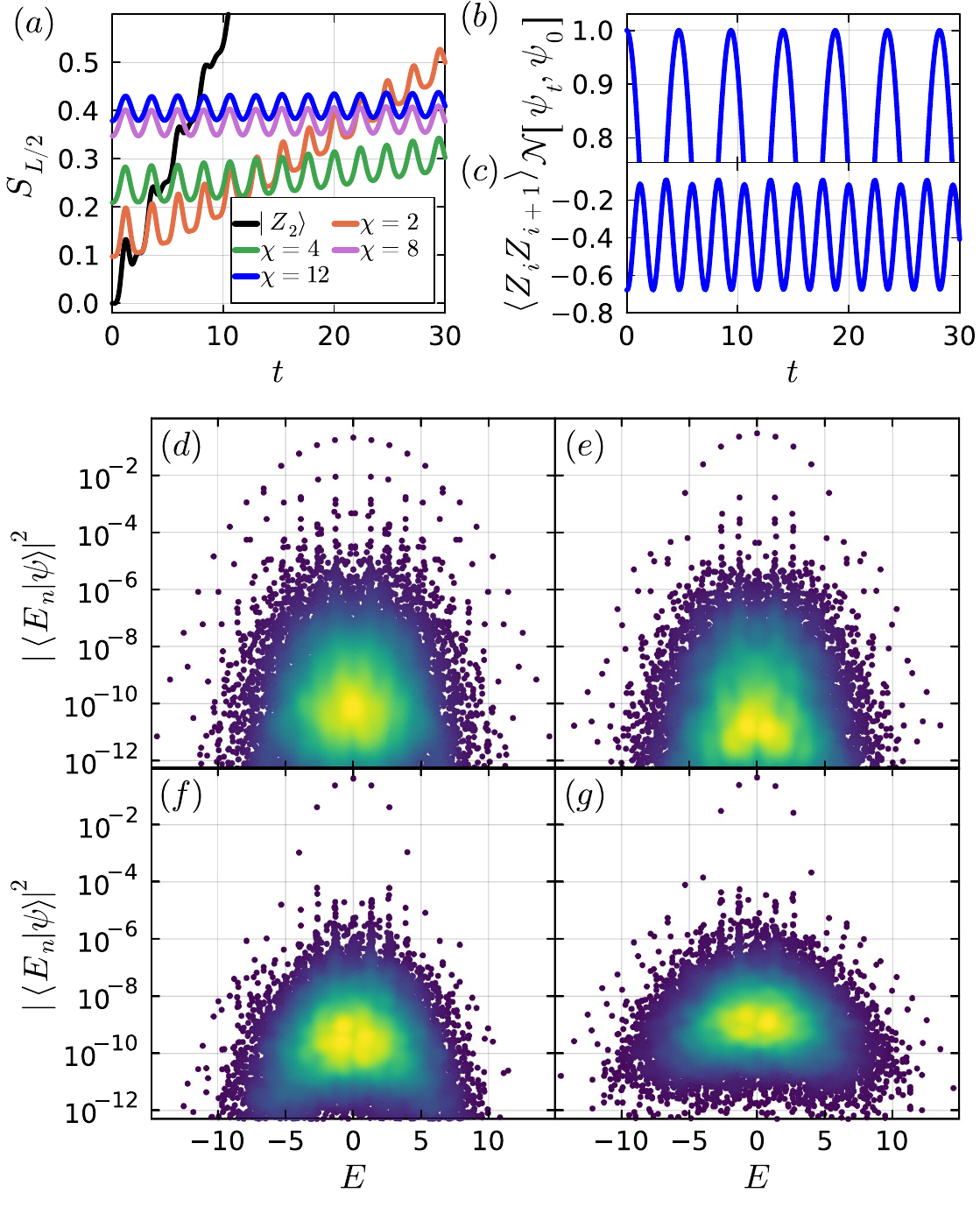}
    \caption{
    (a)-(c) Dynamics of the optimal scar initial state for the PXP Hamiltonian with a two-site unit cell and various bond dimensions \(\chi\).
    (a) Entanglement entropy growth. Compared to the \( |\mathbb{Z}_2\rangle \) state (black line), \(\chi=2\) already shows significantly reduced entanglement growth, while \(\chi=8,12\) dynamics are nearly flat, indicating an extremely stable scar.
    (b) Logarithmic fidelity dynamics for the optimal state at \(\chi=12\), shows nearly perfect revivals with no visible decay.
    (c) Expectation value of the nearest-neighbor observable \( \langle Z_i Z_{i+1} \rangle \) for \(\chi=12\), showing persistent oscillations throughout the evolution.
    (d)-(g) Overlap between the optimized scar state \( |\psi(\chi)\rangle \) and the eigenstates of the PXP Hamiltonian for different bond dimensions:
    (d) \(\chi=2\), (e) \(\chi=4\), (f) \(\chi=8\), and (g) \(\chi=12\).
    As \(\chi\) increases, a well-defined band of scarred eigenstates becomes more distinct from the thermal bulk, with increasing spectral isolation.
    Data in panels (a)-(c) are obtained by iTEBD, while (d)-(g) are exact diagonalization results for system size \(L=24\) in \(k=0\) momentum sector. Since iTEBD simulations in (a)-(c) use a 4-site unit cell, the momentum $k=0$ in exact diagonalization data corresponds to translation by 4 sites.
    }
    \label{fig:pxp_k2_ee_fed}
\end{figure}

We run \textsc{ScarFinder} on $\hat H_\mathrm{PXP}'$ by sampling 100 random initial states in $\mathcal{M}$ and selecting one that exhibits the lowest entanglement entropy at time \(t = 30\).\footnote{We use the iTEBD algorithm to simulate the time evolution. Since local terms in the PXP Hamiltonian act on three neighboring sites, a four-site unit cell is required to implement the Trotter decomposition. Nevertheless, the resulting dynamics largely preserve the underlying \emph{two-site} translational symmetry.} The resulting optimal dynamics are shown in Fig.~\ref{fig:pxp_k2_ee_fed}(a)-(c). Notably, we find that even with a relatively small bond dimension \(\chi = 8\), the algorithm produces an initial state that shows minimal entanglement growth and remarkably robust fidelity revivals—signatures of scar dynamics that persist over long times (\(t \lesssim 30\)) in the thermodynamic limit. We remark that the results presented here are based on minimal entanglement entropy criterion; however, they are not the only scar solutions identified by \textsc{ScarFinder}. 
Specifically, for $\chi=3$ (and also $\chi=4$), we also find the exact $E=0$ MPS eigenstate of the PXP Hamiltonian reported in Ref.~\cite{lin2019exact}.

In order to understand the mechanism for the dynamics observed in Fig.~\ref{fig:pxp_k2_ee_fed}(a)-(c), it is useful to inspect the eigenstate composition of the optimal scar state. As shown in Fig.~\ref{fig:pxp_k2_ee_fed}(d)-(g), the states belonging to the scar band generally split into two subsets as $\chi$ is increased. The subset closer to the middle of spectrum develops an enhanced overlap with the optimal scar state, while the overlaps of the other subset diminish and approach those of typical eigenstates in the bulk of the spectrum.  Thus, an 
increase in $\chi$ results in a clearer separation between some of the scarred eigenstates and the other eigenstates. Importantly, we find that these high-overlap eigenstates coincide with those observed in the \( |\mathbb{Z}_2\rangle \) case. Thus, an increase in $\chi$ effectively tunes the coefficients of these scar eigenstates and picks out a superposition that approximates better and better the optimal scar state. A qualitatively similar enhancement of overlaps near the middle of the spectrum was observed in Ref.~\cite{PhysRevB.102.195114}, albeit in a different setting where the PXP Hamiltonian was perturbed to enhance the scar dynamics.

 \subsection{Three-site unit cell and energy conservation}\label{sec:3site}

Beyond the well-known two-site unit cell dynamics, the PXP model also supports approximate periodic revivals on a three-site unit cell, when the system is prepared in the \( |\mathbb{Z}_3\rangle \equiv \left|\uparrow\downarrow\downarrow\uparrow\downarrow\downarrow \cdots \right\rangle \) state~\cite{turner2018weak,Turner2018PRB}. These revivals, however, are much weaker than those associated with the $|\mathbb{Z}_2\rangle$ initial state. Ref.~\cite{Michailidis2020} constructed an MPS state with \( \chi = 2 \) that showed improved revival fidelity compared to the simple \( |\mathbb{Z}_3\rangle \) product state. We now apply the \textsc{ScarFinder} algorithm to the three-site unit cell in order to systematically explore how well the $|\mathbb{Z}_3\rangle$ revivals can be improved by increasing the amount of entanglement in the initial state.

The procedure is identical to the two-site case, except for adapting the iMPS ansatz to a three-site periodic structure. However, during the random sampling of initial states, we often find states whose dynamics are governed by the same set of eigenstates identified in the two-site unit cell setting. Thus, the $\mathbb{Z}_2$ orbit acts as an attractor for the three-site dynamics.
The appearance of \(\mathbb{Z}_2\)-type oscillations in the three-site unit cell setting can be understood as follows: the $|\mathbb{Z}_2\rangle$ initial state has nonzero overlap with eigenstates in the momentum \(k=0\) sector defined under one-site translation symmetry. This \(k=0\) sector is fully contained within the \(k=0\) sector defined under three-site translation symmetry. As a result, we naturally expect to observe remnants of the \(\mathbb{Z}_2\)-type scar dynamics even when working with a three-site unit cell. To isolate genuine three-site scar trajectories, we perform a post-selection based on the oscillation frequency, allowing us to filter out spurious results dominated by enhanced overlaps with these residual \(\mathbb{Z}_2\) modes. These are easily distinguished by their revival period: the intrinsic three-site scar dynamics feature a shorter oscillation period compared to those from the two-site case~\cite{Turner2018PRB}. 

\begin{figure}[tb]
    \centering
    \includegraphics[width=\linewidth]{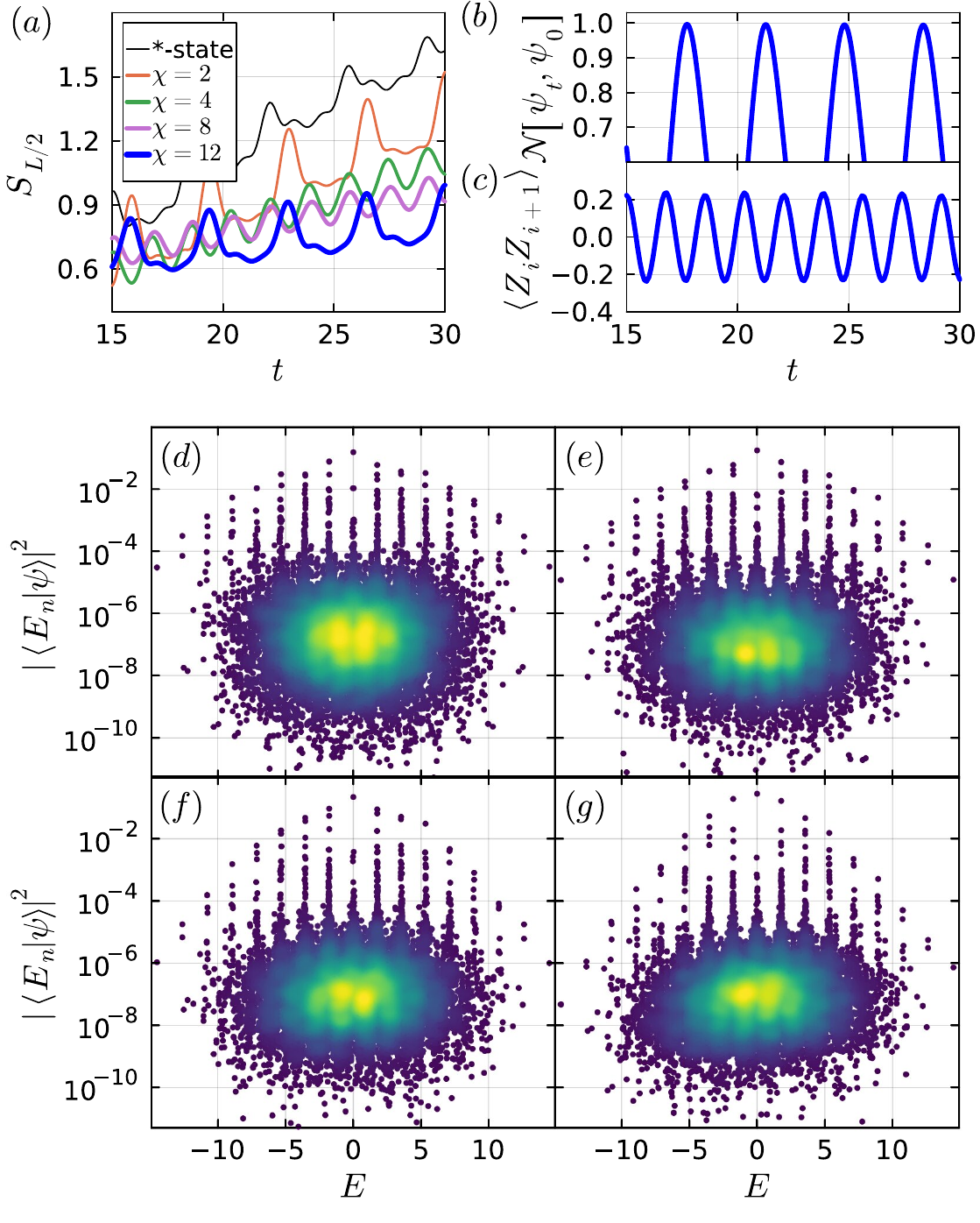}
    \caption{
    (a)-(c) Dynamics of the optimized scar initial states for the PXP Hamiltonian with a 3-site unit cell and different bond dimensions \(\chi\).
    (a) Late-time entanglement entropy growth (\(t \in [15,30]\)). Increasing \(\chi\) systematically suppresses the entanglement growth and stabilizes periodic dynamics. The ``*-state'' refers to the MPS-optimized initial state from Ref.~\cite{Michailidis2020}, previously shown to have better revivals compared to the \(|\mathbb{Z}_3\rangle\) product state.
    (b) Late-time (\(t \in [15,30]\)) logarithmic fidelity for the optimized initial state at \(\chi=12\), computed from the dominant eigenvalue of the MPS transfer matrix. The fidelity displays persistent revivals, highlighting the stability of the identified scar trajectory.
    (c) Time evolution of the nearest-neighbor observable \( \langle Z_i Z_{i+1} \rangle \) for \(\chi=12\).
    (d)-(g) Overlaps between energy eigenstates and the optimized initial scar state for bond dimensions
    \(\chi=2\) (d), \(\chi=4\) (e), \(\chi=8\) (f), and \(\chi=12\) (g). 
    With increasing \(\chi\), distinct eigenstate towers clearly emerge from the thermal continuum, signifying improved scar dynamics.
    Data in panels (d)-(g) is obtained by exact diagonalization for a system size \(L=24\) in the \(k=0\) momentum sector w.r.t. 3-site translations.
    }
    \label{fig:pxp_k3_eefed}
\end{figure}

The resulting optimal trajectories in a 3-site unit cell are shown in Fig.~\ref{fig:pxp_k3_eefed}(a). For comparison, the so-called ``*-state" introduced in Ref.~\cite{Michailidis2020} was constructed using a fixed MPS ansatz with bond dimension \( \chi = 2 \). We observe that with increasing bond dimension, the entanglement growth of the optimized scar states is further reduced, and the fidelity revivals become more pronounced. 
For $\chi=12$ [Fig.~\ref{fig:pxp_k3_eefed}(b)], the fidelity revival is nearly perfect, although not as high as in the two-site unit cell case. In addition, the improved scar dynamics is further supported by the oscillatory behavior of the nearest-neighbor observable \( \langle Z_i Z_{i+1} \rangle \), shown in Fig.~\ref{fig:pxp_k3_eefed}(c).

In Fig.~\ref{fig:pxp_k3_eefed}(d)-(g), we also plot the overlaps of the initial scar states with the energy eigenstates. Interestingly, unlike the two-site unit cell case where the overlap distribution is more dispersed, the three-site unit cell dynamics produce discrete towers of states with approximately equal energy spacing, highlighting a distinct and more structured scar subspace.

\section{Identifying stable islands in Poincar\'e sections}\label{sec:poincare}

In few-body dynamical systems, periodic dynamics and atypical eigenstates can emerge from the coexistence of regular and chaotic regions, resulting from  weakly-broken integrability described by the Kolmogorov-Arnold-Moser (KAM) theorem~\cite{ArnoldBook,GutzwillerBook}. This phenomenon is known as \emph{mixed phase space} and it has many physical realizations, including the Chirikov map, the Fermi-Pasta-Ulam system of coupled oscillators, the Lorentz attractor, and many other~\cite{stockmann2000quantum}.  
For a general dynamical system, mixed phase space is diagnosed by recording successive intersections of trajectories with a chosen lower-dimensional transverse surface in phase space -- a map known as the Poincar\'e section~\cite{strogatz2001nonlinear}. Periodic trajectories correspond to stationary
points of the Poincar\'e map, while during the chaotic dynamics, the system returns to the same surface at a location that is generally far away from a previous encounter. 

Ref.~\cite{Michailidis2020} demonstrated that analogues of Poincar\'e sections can be identified in \emph{many-body} quantum systems, once their dynamics is projected to a low-dimensional MPS manifold. The resulting TDVP equations of motion within the manifold then serve as an effective mapping of the many-body quantum system onto a few-body classical (non-linear) dynamical system. For the latter, standard tools~\cite{strogatz2001nonlinear} can be employed to obtain Poincar\'e sections. However, such analyses become increasingly cumbersome in higher-dimensional manifolds. A pertinent practical question is: can one directly identify the centers of KAM tori associated with scar dynamics?

From the above description of the Poincar\'e section, it should  be clear that it has many things in common with the \textsc{ScarFinder} algorithm. As discussed in Sec.~\ref{sec:algorithm}, the \textsc{ScarFinder} algorithm can be viewed as generating discrete-time dynamics on the variational manifold. Although the theoretical justification of \textsc{ScarFinder} naturally favors larger projection time steps \(\Delta t\), the method remains well-defined in the continuous-time limit \(\Delta t \rightarrow 0\). In this limit, \textsc{ScarFinder} dynamics become equivalent to the TDVP evolution equations~\cite{Michailidis2020}. This connection implies that, for a given MPS ansatz describing the manifold $\mathcal{M}$,  the continuous-time limit \(\Delta t \rightarrow 0\) of the \textsc{ScarFinder} algorithm should recover precisely the TDVP dynamics within $\mathcal{M}$. 
Since any numerical implementation of \textsc{ScarFinder} necessarily uses finite time steps \(\Delta t\),  such discretizations will inevitably introduce small, systematic deviations from the ideal TDVP trajectory. Surprisingly, rather than causing the dynamics to become chaotic, these deviations accumulate constructively, driving the trajectory towards stable fixed points—precisely the centers of stable islands in the mixed phase space portrait. 
This motivates the following  generalization of the \textsc{ScarFinder} algorithm for exploring Poinca\'re sections.

\subsection{\textsc{ScarFinder} algorithm for stable islands in Poincar\'e sections}\label{app:poincare}

We propose the following protocol to directly identify stable islands in a Poincar\'e section via the following modification of \textsc{ScarFinder} algorithm:
\begin{algorithm}[H]
\caption{Poincar\'e \textsc{ScarFinder} algorithm}
\label{alg:findpoincare}
{\bf Input:} An appropriate manifold $\mathcal{M}$, Hamiltonian $\hat{H}$, projection time step $\Delta t$, number of iteration steps $N_\text{step}$,  Poincar\'e sampling number $N_\text{sample}$. \\
{\bf Output:} Poincar\'e section data points
{\setstretch{1.35}
\begin{algorithmic}[1]
\For{$n=1:N_\text{sample}$}
\State \hangindent1.5em\hangafter1 {\bf Initialize:} Initialize an MPS state $|\psi(A(\bm{\theta}))\rangle$ with matrices $A(\bm{\theta})$ and randomly chosen parameters  $\bm{\theta} = (\theta_1, \theta_2, \theta_3,\ldots)$; 
\For{$m=1:N_\text{step}$}
\State Evolve $|\psi\rangle$ to $|\psi(\Delta )\rangle \coloneq \exp(-i\hat{H}\Delta t)|\psi\rangle$; 
\State\label{alg:poincare_project} \hangindent3em\hangafter1 Project $|\psi(\Delta t)\rangle$ back onto $\mathcal M$ by maximizing the logarithmic overlap, $\max_{A({\bm\theta})}\mathcal{N}[\psi^\prime, \psi(A(\bm{\theta}))]$, $|\psi'\rangle \coloneq \hat{P}_\mathcal{M}|\psi(\Delta t)\rangle$;
\If{The trajectory in parameter space crosses the plane of interest}
\State \hangindent4.5em\hangafter1Record the intersection point obtained by linear interpolation between consecutive points;
\EndIf
\State Update $|\psi\rangle \coloneq |\psi^\prime\rangle$.
\EndFor
\EndFor
\end{algorithmic}
}
\end{algorithm}

We note that, in general, the projection step (line~\ref{alg:poincare_project}) should include an energy correction to ensure that the target energy density is preserved. This can be achieved by slightly modifying the projection procedure to incorporate the energy constraint during the maximization. However, for certain models such as the PXP model studied in Sec.~\ref{sec:pxp_mixed} below, this correction turns out to be unnecessary, as the chosen variational manifold inherently preserves the energy.

\subsection{Application to PXP mixed phase space}\label{sec:pxp_mixed}

\begin{figure}[tb]
  \centering
  \includegraphics[width=\linewidth]{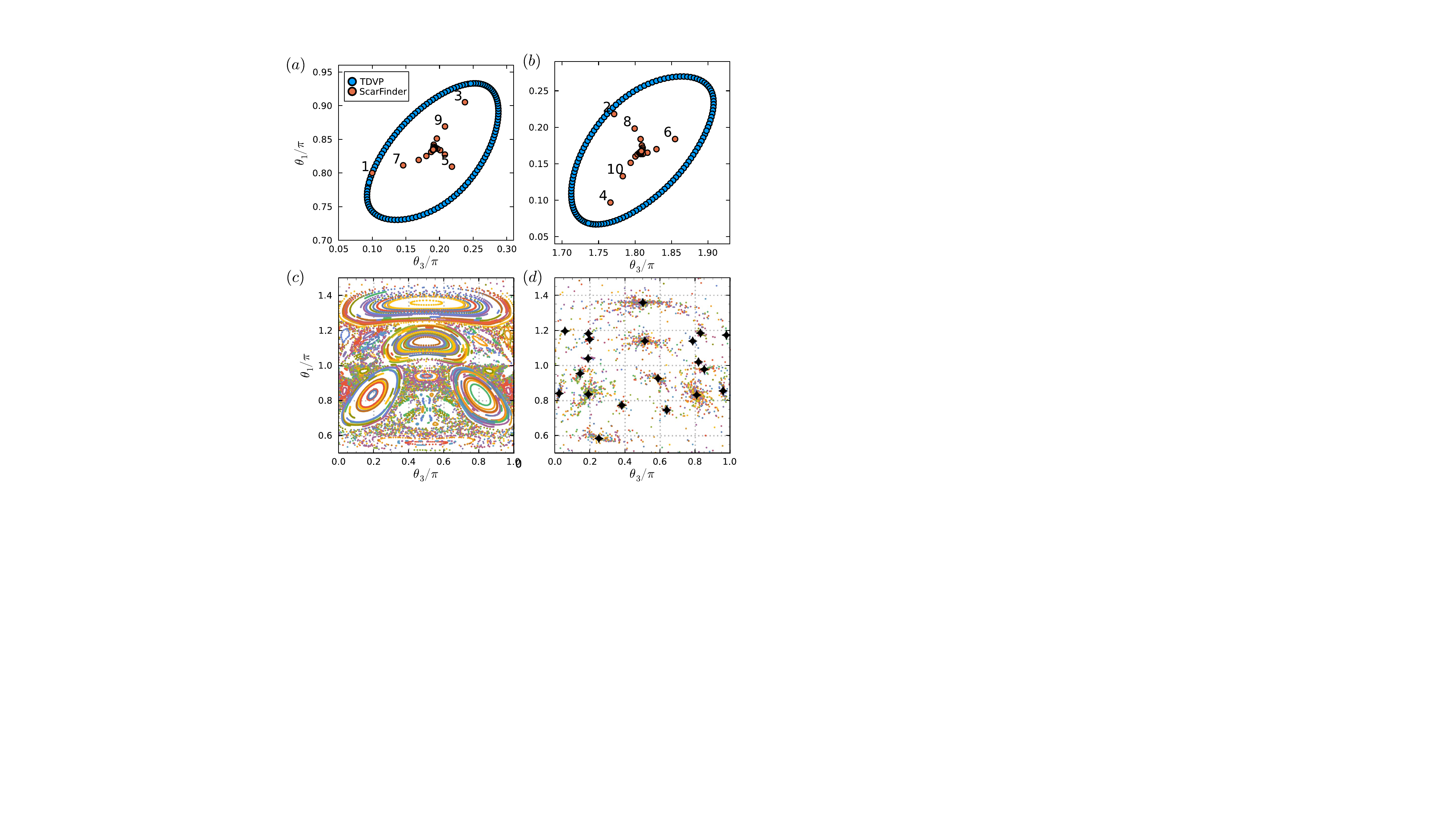}
  \caption{(a),(b) Comparison of Poincar\'e sections obtained via TDVP and \textsc{ScarFinder}. Trajectories obtained from TDVP (blue points) and \textsc{ScarFinder} (orange points with numbers indicating iteration steps), starting from the same initial condition. The trajectory under \textsc{ScarFinder} spirals rapidly towards the center of the stable island, corresponding exactly to the robust scar trajectory identified as the ``*-state" in Ref.~\cite{Michailidis2020}. The figure is divided into two panels, (a) and (b), for clarity, showing two distinct regions where the trajectories cross the plane.
  (c) Our reproduction of the TDVP Poincar\'e section from Ref.~\cite{Michailidis2020}, displaying a mixture of stable islands and chaotic regions. 
  (d) \textsc{ScarFinder} directly identifies the centers of stable islands in (c).}
  \label{fig:pxp_section}
\end{figure}

Let us illustrate the Poincar\'e-\textsc{ScarFinder} algorithm on the example considered in Ref.~\cite{Michailidis2020}. We assume the PXP model with a simple two-dimensional manifold defined by the following MPS ansatz, first introduced in Ref.~\cite{Ho2019}:
\begin{equation}\label{eq:MPS}
  A^{\uparrow}(\theta_i) = \begin{pmatrix}
    0 & i \\
    0 & 0
  \end{pmatrix}, \quad
  A^{\downarrow}(\theta_i) = \begin{pmatrix}
    \cos \theta_i & 0 \\
    \sin \theta_i & 0
  \end{pmatrix}.
\end{equation}
Like in Sec.~\ref{sec:3site}, we assume 3-site periodicity with three MPS angles, $\theta_1$, $\theta_2$ and $\theta_3$, which is the minimal manifold that allows for quantum chaos.  We initialize the angles to \(\bm{\theta}=(0.8,0,0.1)\pi\), and compare the evolution under standard TDVP dynamics with the discrete-time \textsc{ScarFinder} iteration with \(\Delta t = 0.1\). We track the parameter-space trajectory's intersection points with the plane defined by  \(\theta_2=0\). 

Figures \ref{fig:pxp_section}(a) and \ref{fig:pxp_section}(b) illustrate the  contrast between TDVP and \textsc{ScarFinder} dynamics. While TDVP trajectories trace invariant tori, the finite-step \textsc{ScarFinder} trajectory spirals to a unique attractor, located precisely at the center of a stable island. In Fig.~\ref{fig:pxp_section}(c)-(d), we present additional examples, confirming that \textsc{ScarFinder} systematically identifies the main stable-island centers, provided that a sufficient number of initial conditions are sampled.

\begin{figure}[tb]
    \centering
    \includegraphics[width=\linewidth]{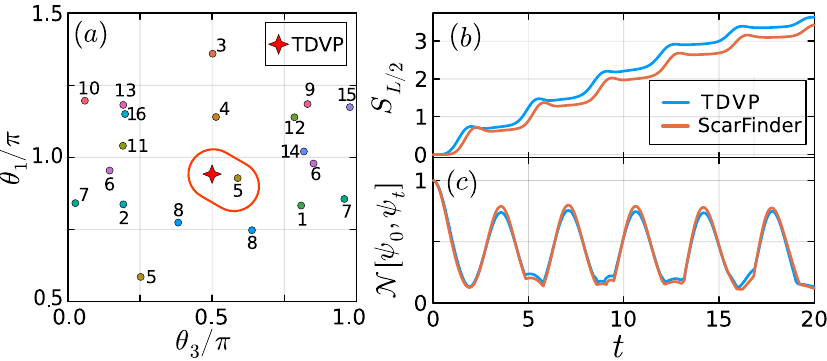}
    \caption{
        (a) Distribution of 16 distinct fixed-point solutions identified by \textsc{ScarFinder} from a total of 640 random initial states within the MPS manifold given by Eq.~(\ref{eq:MPS}). We label the fixed-point sets according to the frequency of getting that solution. The red ellipse highlights two points obtained via different methods: the red star is the optimized state from TDVP-Poincar\'e section, while point 5 was found by ScarFinder.
        (b) Comparison of bipartite entanglement entropy dynamics starting from optimized initial states obtained via TDVP Poincar\'e section [blue line, for the initial state marked by the red star in (a)] and \textsc{ScarFinder} [orange line, for the initial state "5" in (a)].
        (c) Fidelity revival dynamics for the two optimized scar initial states. The state found by \textsc{ScarFinder} exhibits slightly enhanced revivals.
    }
    \label{fig:fixpoint}
\end{figure}

For smaller islands, we observe minor discrepancies between the locations of fixed points found by \textsc{ScarFinder} and the island centers identified from TDVP-generated Poincar\'e sections. To analyze these differences, we extensively sample 640 random initial states within the three-site unit-cell MPS manifold for the PXP model, Eq.~(\ref{eq:MPS}), and identify a total of 16 distinct fixed-point solutions (within $\theta_1 \in [0.5\pi,1.5\pi],\ \theta_3\in[0,\pi]$) using the \textsc{ScarFinder} algorithm, as shown in Fig.\ref{fig:fixpoint}(a). We find that the fixed points labeled ``1'' and ``2'' correspond exactly (within numerical precision) to the ``*-state'' previously discovered in Ref.~\cite{Michailidis2020}. These points coincide precisely with the centers of the largest stable islands identified from the TDVP-generated Poincar\'e sections.

In contrast, the fixed-point set labeled ``5'' [inside the red ellipse in Fig.~\ref{fig:fixpoint}(a)] identified by \textsc{ScarFinder} is located at $\bm\theta_\mathrm{SF} {=} (0.928,\,0,\,0.588)\pi$,
while the corresponding stable island center from the TDVP-generated Poincar\'e section is at: $\bm\theta_\mathrm{TDVP} {=} (0.942,\, 0,\, 0.5)\pi$.
To evaluate the impact of these differences, in Fig.~\ref{fig:fixpoint}(b)-(c) we explicitly compare the dynamics starting from these two initial states. While the dynamical behaviors are qualitatively similar, we observe that the initial state identified by \textsc{ScarFinder} consistently exhibits lower entanglement entropy and noticeably enhanced fidelity revivals.

We thus conclude that the \textsc{ScarFinder} algorithm not only efficiently locates optimal scar states, but in certain cases, even provides better initial conditions compared to states obtained from direct exploration of TDVP-generated mixed-phase spaces. This can be understood by noting that TDVP restricts quantum dynamics to a particular variational ansatz, unlike \textsc{ScarFinder} which utilizes information from exact quantum dynamics during each iteration. The deviation between TDVP and the exact quantum dynamics is characterized using ``quantum leakage''~\cite{Choi2019}. Since scarred trajectories are typically found in regions of small leakage, there is a strong correspondence between their TDVP trajectory and exact quantum dynamics. However, even in these cases the leakage is non-zero and some relevant information has been lost, therefore it is unsurprising that \textsc{ScarFinder}, can improve upon these results as it does not suffer from any leakage effects.

\section{Conclusions}\label{sec:conc}

In this paper we have introduced the \textsc{ScarFinder}:  a general-purpose algorithm for identifying initial conditions that give rise to regular  dynamics in chaotic quantum systems. Using the spin-1 XY model with exact QMBS states, we have performed extensive benchmarks and elucidated the \textsc{ScarFinder} convergence mechanism: the trade-off between the distance of the initial state from the scar trajectory, and the projection time step $\Delta t$ allowing the dynamics to leave the variational manifold $\mathcal{M}$. In particular, we have shown that larger $\Delta t$ ensures faster convergence, even for states initially far from the sought trajectory. We have also pointed out the important role of symmetries and energy conservation. 

Beyond reproducing the known QMBS phenomenology, our application of \textsc{ScarFinder} to the PXP model gave several intriguing new results. In particular, the observations in Fig.~\ref{fig:pxp_k2_ee_fed} and Fig.~\ref{fig:pxp_k3_eefed} hint at an equal energy spacing between (some) scar eigenstates in the thermodynamic limit. Previously, the scar dynamics observed from the \( |\mathbb{Z}_2\rangle \) state were attributed to an \emph{approximately} equal energy spacing among these eigenstates~\cite{turner2018weak,Choi2019}. However, since the \( |\mathbb{Z}_2\rangle \) state does not exhibit perfect revivals and exact diagonalization is limited by finite-size effects, it has remained unclear whether the equal energy spacing persists in the thermodynamic limit. Our improved superposition state in the form of an iMPS and the trend in Figs.~\ref{fig:pxp_k2_ee_fed}-\ref{fig:pxp_k3_eefed} suggest that equal energy spacing is an exact property of some eigenstates in the middle of the spectrum of the PXP model in the thermodynamic limit. This is in line with recent work~\cite{ivanov2025exactarealawscareigenstates} which found several exact MPS eigenstates in the spectrum of the PXP model. 

Although our discussion primarily focused on 1D systems, the core methodology of \textsc{ScarFinder} generalizes naturally to higher dimensions. In Appendix~\ref{app:q1d}, we explore its application to quasi-1D settings, including the PXP model on square and triangular lattices wrapped around a finite cylinder. 
These results demonstrate the versatility of \textsc{ScarFinder} beyond 1D, and for targeting simultaneously the scar initial states as well as scar eigenstates.  For isotropic 2D systems, one could consider extending the approach using projected-entangled-pair state representations~\cite{CiracRMP}, which would involve a significant computational overhead and we  therefore leave it as a future direction. 

We emphasize that the utility of \textsc{ScarFinder} is not limited to QMBS models and it can also aid the identification and understanding of other types of atypical dynamics in chaotic models. For example, in the paradigmatic quantum Ising model in the presence of both transverse and longitudinal fields, Ref.~\cite{Banuls2011} found certain product states that undergo ``weak'' thermalization in contrast to the majority of other initial states that strongly thermalize. While these states were subsequently related to low-lying quasiparticle excitations~\cite{LinMotrunich2017}, their existence is nonetheless surprising given that the model is believed to be ``fully'' chaotic and obey the ETH. In Appendix~\ref{app:mfi}, we show that \textsc{ScarFinder} successfully captures these states and allows to ``refine'' them by including entanglement. More importantly, \textsc{ScarFinder}  allows to search for weakly-thermalizing states, like the ones in Ref.~\cite{Banuls2011}, systematically and in general models, without prior knowledge about their underlying physics.

We conclude by highlighting several advantages of \textsc{ScarFinder}. In 1D systems, the algorithm is numerically efficient compared to exact diagonalization due to its straightforward implementation using MPS. The latter allows \textsc{ScarFinder} to  operate directly in the thermodynamic limit, minimizing finite-size effects and providing direct access to stable, infinite-system scar trajectories.  On the other hand, DMRG-based searches have recently been adapted to target low-entanglement excited eigenstates in finite systems~\cite{Zhang2023DMRGS}. However, this approach does not automatically yield the initial conditions for probing such eigenstates in experiment. As mentioned above, \textsc{ScarFinder} can identify both scar eigenstates and scar initial states, thereby circumventing the challenge of DMRG-type methods when scar eigenstates possess entanglement beyond area law. We note, however, that some volume-law states, such as the rainbow states from Refs.~\cite{PhysRevB.105.L060301,Mohapatra2023}, can still be amenable to \textsc{ScarFinder} with a suitably modified unit cell of the lattice, which minimizes entanglement. Additionally, we have demonstrated that \textsc{ScarFinder} directly identifies the centers of stable islands in Poincar\'e sections, eliminating the need for complicated visualizations of high-dimensional phase spaces. These advantages make \textsc{ScarFinder} a versatile tool for identifying non-thermal behaviors in many-body systems, in particular in the field of quantum simulation and designing long-lived quantum states.

{\sl Note Added:} During the completion of this work, we became aware of a related study by Petrova \emph{et al.}~\cite{petrova2025findingperiodicorbitsprojected}, where a different method based on TDVP and gradient-descent was developed to identify periodic trajectories in the Floquet Ising model. 

\begin{acknowledgments}

We thank Jean-Yves Desaules for useful comments.
J.R., A.H. and Z.P. acknowledge support by the Leverhulme Trust Research Leadership Award RL-2019-015 and EPSRC Grant EP/Z533634/1. 
This research was supported in part by grant NSF PHY-2309135 to the Kavli Institute for Theoretical Physics (KITP). 
L.Y. acknowledges support from the Zhejiang Provincial Natural Science Foundation of China (No. LD25A050002), the National Natural Science Foundation of China (Nos. 12247101 and 12375021), the Fundamental Research Funds for the Central Universities (Grant No. lzujbky-2024-jdzx06), and the National Key Research and Development Program of China (No. 2022YFA1404203).
\end{acknowledgments}

\begin{appendix}

\section{Construction of parent Hamiltonians}
\label{apx:parent-ham}

For a given set of target states $\{|\psi_n\rangle\}$, general frameworks exist for constructing parent Hamiltonians, namely, the \textit{quantum inverse method}~\cite{inverse-method,inverse-method-2} and the \textit{projective embedding method}~\cite{ShiraishiMori}. For the Type-I scar tower in the spin-1 XY model, we adopt the quantum inverse method, which provides an explicit and interpretable Hamiltonian expression. In contrast, for the Type-II scar tower, the resulting Hamiltonians from the inverse method become overly complicated and obscure the underlying physics. In this case, we instead employ the projective embedding method, which yields a structurally simple and formally well-defined Hamiltonian tailored to the desired subspace.

\subsection{Type-I scar tower}\label{app:typeI}

The key object in the quantum inverse method is the \textit{quantum covariance matrix} \(C_T\)~\cite{inverse-method,inverse-method-2}, defined on the target subspace:
\begin{equation}\label{apxeq:qcm}
	(C_\text{target})_{ab}=\frac{1}{2}\langle[\hat h_a,\hat h_b]_+\rangle_\text{target}-\langle\hat h_a\rangle_\text{target}\langle\hat h_b\rangle_\text{target},
\end{equation}
where $[,]_+$ denotes the anticommutator and the set $\{\hat h_a\}$ is the given list of local operators as the building blocks for the parent Hamiltonian, and \(\langle \cdot \rangle_\text{target}\) indicates expectation values averaged over all states in the target subspace: \(\langle \hat{O}\rangle_\text{target} = \frac{1}{N}\sum_{n=1}^N\langle\psi_n|\hat{O}|\psi_n\rangle\).
For the Type-I scar tower, we start by selecting an operator basis consisting of nearest-neighbor couplings for a spin-1 system. Specifically, we consider two-site interactions generated by the following local operators:
    $\{\hat I,\, \hat S^x,\, \hat S^y,\, \hat S^z,\, \hat P^0\}$, where the operator \(\hat{P}^0 = |0\rangle\langle 0|\) projects a local spin onto the zero state.

We numerically construct the covariance matrix \( C_T \) [Eq.~(\ref{apxeq:qcm})] using exact Type-I scar eigenstates as input for a finite system of length \( L=8 \). Explicitly, the scar eigenstates used in this construction are given by Eq.~(\ref{eq:Type1QMBS}).
By computing numerically the null space of \(C_T\), we identify a set of 9 linearly independent Hamiltonian terms under which the given scar states form an exact degenerate eigenspace. These Hamiltonians take the forms:
\begin{eqnarray}
    \nonumber \hat H_0 &=& \sum_j \left(\hat S^x_j \hat S^x_{j+1} + \hat S^y_j \hat S^y_{j+1}\right), \\
    \nonumber \hat H_1 &=& \sum_j\hat P^0_j, \quad
    \hat H_2 = \sum_j\hat P^0_j \hat P^0_{j+1}, \\ \label{eq:apx-xy1-ham}
    \hat H_3 &=& \sum_j\hat P^0_j \hat S^x_{j+1}, \quad
    \hat H_4 = \sum_j\hat S^x_j \hat P^0_{j+1}, \\
    \nonumber \hat H_5 &=& \sum_j\hat P^0_j \hat S^y_{j+1}, \quad
    \hat H_6 = \sum_j\hat S^y_j \hat P^0_{j+1}, \\
    \nonumber \hat H_7 &=& \sum_j\hat P^0_j \hat S^z_{j+1}, \quad
    \hat H_8 = \sum_j\hat S^z_j \hat P^0_{j+1}.  
\end{eqnarray}
We notice that the term \(\hat{H}_0\) corresponds precisely to the standard XY Hamiltonian, while the remaining terms are variations of the general form:
\begin{equation}
    \hat H^l_i = \sum_j \hat{P}^0_j\hat V^i_{j+1}, \quad
    \hat H^r_i = \sum_j \hat V^i_{j}\hat{P}^0_{j+1},
\end{equation}
where \(\hat{V}^i_j\) denotes arbitrary single-site Hermitian operators. Physically, the presence of the operator \(\hat{P}^0_j\) ensures that these Hamiltonian terms annihilate all scar eigenstates, since the type-I scar states contain no local \(|0\rangle\) components. Thus, these terms effectively embed the target scar subspace as degenerate eigenstates within an otherwise chaotic spectrum.

In Ref.~\cite{Schecter2019}, the additional perturbation included was simply \(\hat H_1\) in Eq.~\eqref{eq:apx-xy1-ham}. In the main text, we focus on a representative Hamiltonian of type \(\hat{H}^l\) with the specific choice \(\hat V_j = \hat S^x_j\). For a more robust and generic construction, we further propose the following random Hermitian operator:
\begin{equation}
    \hat V = \sum_{n=1}^8 \sin(100 n)\,\hat \lambda^{(n)},
\end{equation}
where \(\hat \lambda^{(n)}\) is the \(n\)-th Gell-Mann matrix. This fully eliminates any residual structure and promotes optimal convergence of the \textsc{ScarFinder} algorithm.

\subsection{Type-II scar tower}\label{app:typeII}

For the type-II scar tower, the construction is slightly more involved due to the intrinsic MPS structure of the scar eigenstates. To adequately capture this structure, we enlarge our operator basis to include all possible two-site local coupling terms. Following the same numerical approach with exact diagonalization for a finite system of size \( L=8 \), apart from XY Hamiltonian, we obtain 4 additional solutions:
\begin{eqnarray}
    \nonumber \hat H_1 &=& \sum_j \left[(\hat S^+_j)^2 (\hat S^-_{j+1})^2+(\hat S^-_j)^2 (\hat S^+_{j+1})^2\right], \\
    \nonumber \hat H_2 &=& \sum_j \left[i(\hat S^+_j)^2 (\hat S^-_{j+1})^2-i(\hat S^-_j)^2 (\hat S^+_{j+1})^2\right], \\
    \hat H_3 &=& \sum_j \left[(\hat S^z_j)^2 (\hat S^z_{j+1})^2-\hat S^z_j \hat S^z_{j+1}\right], \\
    \nonumber \hat H_4 &=& \sum_j \left[\hat S^z_{j+1} (\hat S^z_{j+2})^2-(\hat S^z_{j+1})^2 \hat S^z_{j+2}\right].    
\end{eqnarray}
These 4 terms, however, also conserve magnetization. Note that these terms all locally annihilate the scar tower, hence we can construct the 3-site interaction:
\begin{equation}\label{eq:V2}
    \hat V_2 = \sum_j \left[(\hat S^+_j)^2 (\hat S^-_{j+1})^2+(\hat S^-_j)^2 (\hat S^+_{j+1})^2\right] \hat S_{j+1}^x.
\end{equation}
Nevertheless, this term still contains some structure. 
To get a fully thermalizing perturbation, we use the projective embedding method~\cite{ShiraishiMori} on a 3-site cluster:
\begin{equation}\label{apxeq:xy2_embed_ham}
    \hat V_2' = \sum_j(1-\hat P^{[3]})_j \hat h^\text{rand}_j(1-\hat P^{[3]})_j,
\end{equation}
where $\hat P^{[3]}$ is the local projector such that $(1-\hat P^{[3]})$ locally annihilates all scar states.

\section{Pedagogical example of spin-1 XY model}
\label{apx:finite_xy}

In this Appendix, we explore explicitly how $\eta$ in Eq.~(\ref{eq:scar_decomp}) updates during the Schmidt truncation. We restrict ourselves to a sufficiently small system size \( L = 6 \) with periodic boundary condition that can be exactly diagonalized. We fix the Hamiltonian with $\hat V=\hat V_1$, choose \( h = 1 \), and consider the imperfect initial state in Eq.~(\ref{eq:imperfect_scar}), with \( \alpha = \pi/6 \). Numerically, we find that this state has a scar overlap characterized by \( \eta \approx 0.583 \) in the decomposition defined in Eq.~\eqref{eq:scar_decomp}.

\begin{figure}[tb]
    \centering
    \includegraphics[width=\linewidth]{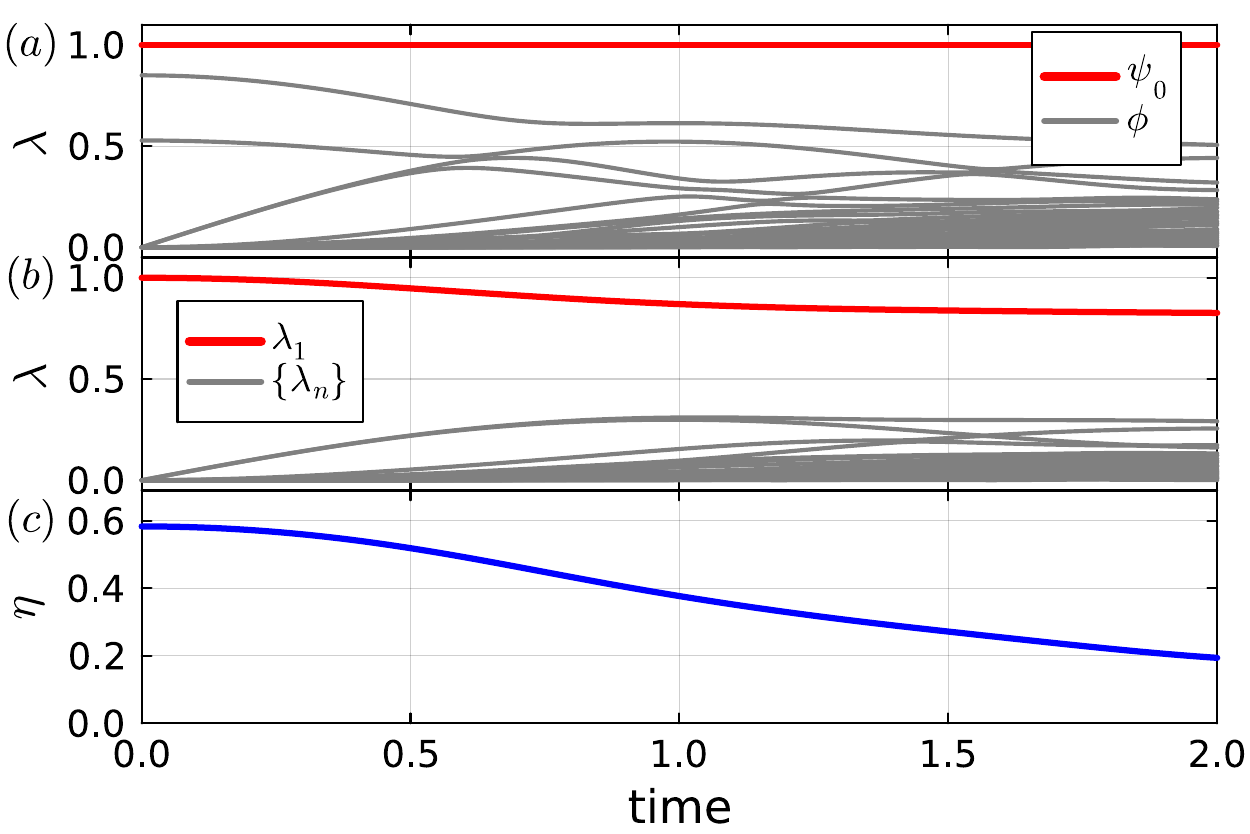}
    \caption{(a) Singular values \( \lambda_k(t) \) of the time-evolved states $\psi_0$ (red lines) and $\phi$ (grey lines). (b) Singular values \( \lambda_k(t) \) of the combined time-evolved state $\psi$, with the dominant Schmidt value plotted in red. (c) The extracted \( \eta \)-value corresponding to the dominant Schmidt component at each instant in time.
    }
    \label{fig:xy1_eta}
\end{figure}

We now consider the Schmidt decomposition of the time-evolved state:
\begin{equation}\label{eq:xy-schmidt}
    e^{-i\hat{H} t} |\psi\rangle = \sum_{k=1}^{27} \lambda_k( t)\, |\psi_A^k\rangle \otimes |\psi_B^k\rangle,
\end{equation}
where \( \lambda_k(t) \) are the singular values. Since the initial state is a product state, only one singular value is non-zero at \( t = 0 \). As time passes, entanglement builds up due to the imperfect initial state, and additional singular values emerge. However, because the scar component dominates, a significant gap between the leading and subleading singular values persists throughout the evolution, as shown in Fig.~\ref{fig:xy1_eta}.

The key insight of the \textsc{ScarFinder} algorithm lies in the observation that the dominant Schmidt component,
\[
    \lambda_1(t)\, |\psi_A^1\rangle \otimes |\psi_B^1\rangle,
\]
contains more scar content than thermal content. Hence, when the variational manifold \( \mathcal{M} \) is defined as the set of all bipartite product states across the half-cut, the projection step becomes a simple truncation to the leading Schmidt term:
\begin{equation}
    \hat{P}_\mathcal{M} |\psi(t)\rangle = \lambda_1(t)\, |\psi_A^1\rangle \otimes |\psi_B^1\rangle.
\end{equation}
Choosing a fixed projection time step $\Delta t$, a single step gives
\[
|\psi\rangle \rightarrow \hat{P}_\mathcal{M}\left[e^{-i\hat H \Delta t}|\psi\rangle\right].
\]
By evaluating the \( \eta \)-value of the projected state, we find that it tends to decrease with time, as shown in Fig.~\ref{fig:xy1_eta}(c).  

We iterate this projection process with a time step of \( \Delta t = 2.0 \). The resulting sequence of \( \lambda \)-values is shown in Table~\ref{tab:xy1_eta}, confirming steady convergence toward the scar trajectory.

\begin{table}[h]
    \centering
    \begin{tabular}{c|ccccccccc}
        \hline\hline
        Step & 0 & 1 & 2 & 3 & 4 & 5 & 6 \\
        \hline
        \( \eta \) & 0.5833 & 0.1938 & 0.0657 & 0.0264 & 0.0107 & 0.0050 & 0.0024 \\
        \hline\hline
    \end{tabular}
    \caption{Convergence of the \textsc{ScarFinder} algorithm. Each entry shows the value of \( \eta \) after successive steps of time evolution (\( \Delta t = 2.0 \)) followed by projection onto the product-state manifold \( \mathcal{M} \).}
    \label{tab:xy1_eta}
\end{table}

\section{Matrix-product-state scar tower for spin-1 XY model}
\label{apx:xy2}

When the perturbation is chosen according to $\hat V \equiv \hat V_2$ in Eq.~(\ref{eq:V2}),  
the spin-1 XY model hosts a different type of scar tower, whose linear combination defines the following exact MPS state:
\begin{equation}\label{eq:xyMPSinitialstate}
\begin{aligned}
	A_j^{[+]} &= \begin{bmatrix}
		\sin \frac{\phi}{2} e^{i\theta} & 0 \\ 0 & 0
	\end{bmatrix}, \quad
	A_j^{[0]} = \begin{bmatrix}
		0 & \cos \frac{\phi}{2} \\ (-1)^j\sin \frac{\phi}{2} & 0
	\end{bmatrix}, \\
	A_j^{[-]} &= \begin{bmatrix}
		0 & 0 \\ 0 & (-1)^j\cos \frac{\phi}{2}e^{-i\theta}
	\end{bmatrix}.
\end{aligned}
\end{equation}
The energy density of this state is set by the parameter \( \phi \) according to $E(\phi) = h \cos\phi$. 

\begin{figure}[tb]
    \centering
    \includegraphics[width=\linewidth]{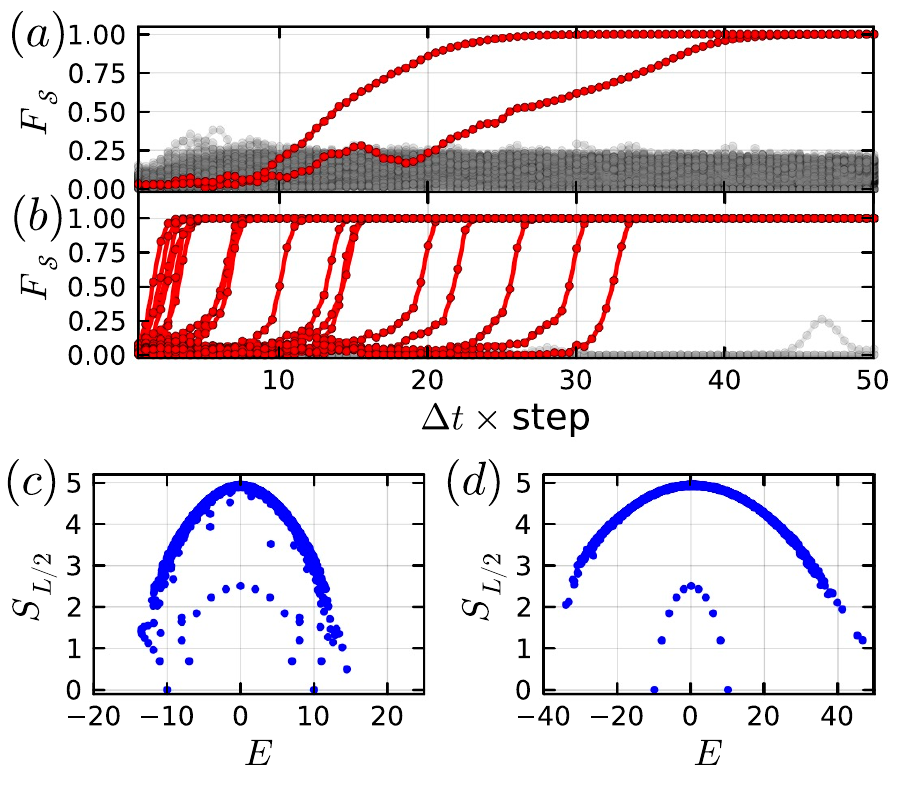}
    \caption{(a)-(b) Convergence behavior of \textsc{ScarFinder} for the spin-1 XY model initialized from random states with energy correction (at target energy \(E_\text{target}=0\)), with $\Delta t=0.5$. 
    (a) For the Hamiltonian with \(\hat V=\hat V_2\) in Eq.~(\ref{eq:V2}), among 200 independent trials, only 2 successfully converge. 
    (b) For the Hamiltonian with \(\hat V = \hat V_2'\) in Eq.~(\ref{apxeq:xy2_embed_ham}), 18 out of 20 trials converge successfully.
    (c)-(d) Bipartite entanglement entropy of eigenstates for the spin-1 XY Hamiltonian [Eq.~(\ref{eq:spin-1XY})] with (c) \( \hat{V} = \hat{V}_2 \) and (d) \( \hat{V} = \hat{V}_2' \). The results are obtained via exact diagonalization on a finite system of size \(L=10\), combining the momentum sectors \(k=0\) and \(k=\pi\).  
    }
    \label{fig:xy2_ee}
\end{figure}

As discussed in Appendix~\ref{apx:parent-ham}, besides $\hat V_2$, we can construct a more generic perturbation $\hat V_2'$ [Eq.~(\ref{apxeq:xy2_embed_ham})]. 
Fig.~\ref{fig:xy2_ee}(a)-(b) compare the convergence behaviour of \textsc{ScarFinder} with these two types of perturbations. 
Fig.~\ref{fig:xy2_ee}(c)-(d) contrast the entanglement entropy spectra for these two perturbations, \( \hat{V}_2 \) and \( \hat{V}_2' \). Compared to panel (c), panel (d) exhibits a clearer separation between low-entanglement scar states and highly-entangled thermal states. This sharper distinction suggests that the modified perturbation \( \hat{V}_2' \) creates a more favorable spectral structure that improves the performance of the \textsc{ScarFinder} algorithm.

\section{PXP-type models on quasi-1D lattices}
\label{app:q1d}

\begin{figure}[t]
    \centering
    \includegraphics[width=\linewidth]{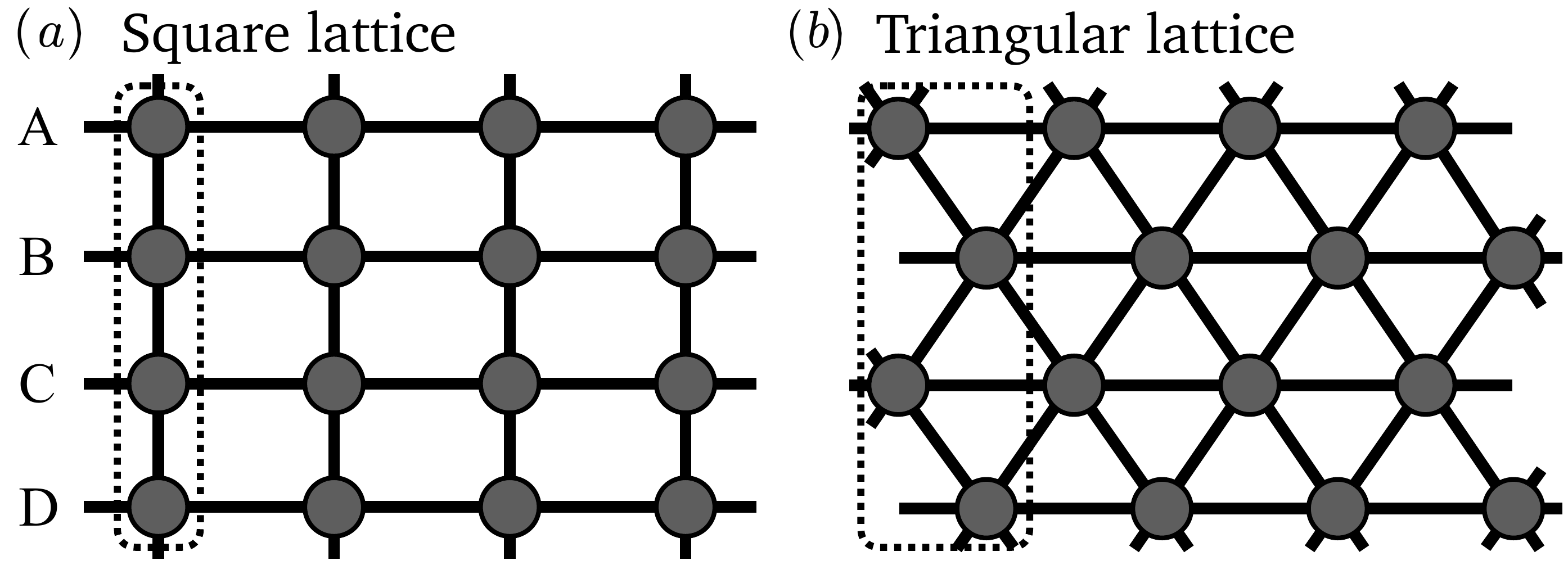}
    \caption{Quasi-1D versions of the PXP model defined on (a) square lattice and (b) triangular lattice. Both systems form cylindrical tubes with periodic boundary conditions along the vertical direction and infinite length horizontally. The Rydberg blockade constraint is imposed along the nearest-neighbor links. Each block, denoted by dashed line, outlines a unit cell that is repeated along the horizontal axis. The unit cell is composed of four sublattice sites labeled \( \mathrm{A}, \mathrm{B}, \mathrm{C}, \mathrm{D} \), which are used to define the MPS ansatz for \textsc{ScarFinder}.}
    \label{fig:Q1DPXP}
\end{figure}

While the main text focuses on 1D models, the core ideas behind \textsc{ScarFinder} are not restricted to that setting. In principle, the algorithm can be extended to higher-dimensional systems, provided one can define a suitable low-entanglement variational manifold. However, such generalizations face significant technical challenges, as entanglement structures are more complex in higher dimensions and tensor network methods are much more difficult to implement~\cite{CiracRMP}.
As a practical intermediate step, we explore quasi-1D  lattice geometries which go beyond simple chains in structure, but retain limited width, allowing for efficient simulation using MPS-based methods. In the following, we apply \textsc{ScarFinder} to the PXP model defined on square and triangular lattice tubes, illustrated in Fig.~\ref{fig:Q1DPXP}. We demonstrate that the algorithm remains effective in uncovering scarred dynamics, while it also allows to find exact eigenstates even in these extended quasi-1D settings.

\begin{figure}[t]
    \centering
    \includegraphics[width=\linewidth]{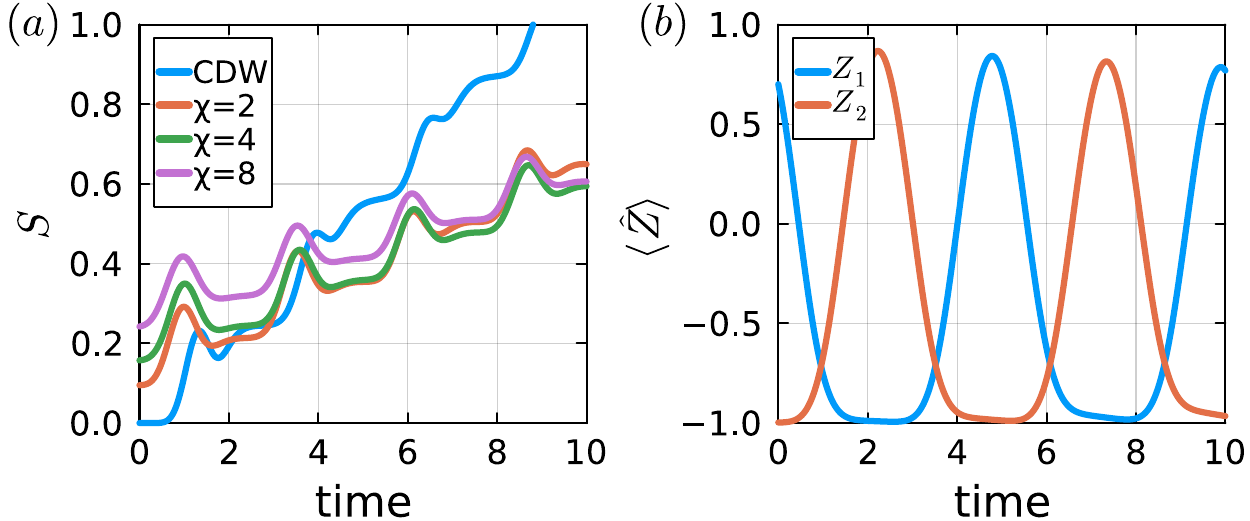}
    \caption{Dynamics of the PXP model on a quasi-1D lattice. (a) Entanglement entropy evolution starting from the charge density wave (CDW) state and from optimized initial states obtained by \textsc{ScarFinder} with various bond dimensions $\chi = 2, 4, 8$. The optimized states exhibit suppressed entanglement growth compared to the CDW state. (b) Local spin expectation values $\langle \hat{Z} \rangle$ on two sublattices show nearly perfect revival dynamics from the optimized state with $\chi = 8$. We ran \textsc{ScarFinder} with projection time step $\Delta t = 1.0$ for 200 iterations and the underlying dynamics were simulated using the iTEBD algorithm with a maximum bond dimension of $\chi_{\text{max}} = 24$.}
    \label{fig:PXPS_SF}
\end{figure}

\subsection{Square lattice on a thin cylinder}

Previously, for the PXP model on a 2D square lattice, Refs.~\cite{lin2020quantum,PhysRevResearch.2.022065} studied the dynamics from a charge-demsity-wave state, $|\text{CDW}\rangle$, with excitations on every site of one of the two sublattices of this bipartite lattice. Qualitatively, the scar dynamics were found to behave similarly as in the 1D case. 
As a first extension, we apply \textsc{ScarFinder} to the PXP model defined on a square lattice, but wrapped onto a finite cylinder shown in Fig.~\ref{fig:Q1DPXP}(a). 
We consider a cylinder defined by periodic boundary conditions along one spatial dimension, while leaving the other dimension infinite. In this quasi-1D setting, the lattice can be described using MPS methods by grouping sites around the circumference into composite sites, as depicted by dashed lines in Fig.~\ref{fig:Q1DPXP}(a).

Specifically, we study the square lattice with circumference $L_y=4$ and infinite length along the $x$-direction. We apply \textsc{ScarFinder} by employing translationally-invariant MPS representation adapted to this geometry, where each physical site represents one circumference unit cell of the cylinder. Note that, because of the constraint,  the local Hilbert space dimension for each blocked site is $d=7$.
For notational simplicity, we define the following 4-site configurations of the block:
\begin{equation}\label{eq:4site}
    |0\rangle \equiv \left| \downarrow\downarrow\downarrow\downarrow\right\rangle,\quad
    |\mathbb{Z}_2\rangle \equiv \left| \uparrow\downarrow\uparrow\downarrow\right\rangle,\quad
    |\mathbb{Z}_2'\rangle \equiv \left| \downarrow\uparrow\downarrow\uparrow\right\rangle.
\end{equation}
The CDW state can be naturally expressed as a product of alternating $|\mathbb{Z}_2\rangle$ and $|\mathbb{Z}_2'\rangle$ configurations: $|\text{CDW}\rangle = |\mathbb{Z}_2,\mathbb{Z}_2',\dots\rangle$.

Now we can apply \textsc{ScarFinder} starting from the $|\text{CDW}\rangle$ state and choosing bond dimensions $\chi=2,4,8$ to find an optimized scar initial state. The resulting entanglement dynamics are shown in Fig.~\ref{fig:PXPS_SF}(a). Compared to the CDW state, which exhibits steadily increasing entanglement the optimized states found by \textsc{ScarFinder} display slower entanglement growth with clearly visible revivals, though sharing the same revival frequency as the $|\text{CDW}\rangle$ state. As the bond dimension increases from $\chi=2$ to $\chi=8$, these revivals become more pronounced, and the entanglement entropies become substantially lower than that of the $|\text{CDW}\rangle$ state at late times.

To further illustrate the coherent dynamics, Fig.~\ref{fig:PXPS_SF}(b) shows the time evolution of local observables $\langle \hat{Z}_1 \rangle$ and $\langle \hat{Z}_2 \rangle$, representing magnetization on two underlying sublattices. The data correspond to dynamics initialized from the optimized state with $\chi=8$ and reveal stable periodic oscillations. These oscillations qualitatively resemble those of the $|\text{CDW}\rangle$ state, but the optimization provided by \textsc{ScarFinder} yields notably improved revival stability and reduced entanglement growth. These results confirm that \textsc{ScarFinder} effectively enhances the scar dynamics of the quasi-1D PXP model defined on a finite cylinder. While numerical limitations currently restrict our analysis to such quasi-1D settings, it would be interesting to verify if a similar optimization procedure could also yield improved scar initial states in the full 2D PXP model.

\subsection{Triangular lattice on a thin cylinder}

As a further application, we apply \textsc{ScarFinder} to the PXP model on a triangular lattice wrapped around a cylinder, with the vertical circumference of four lattice sites and infinite length in the horizontal direction [see Fig.~\ref{fig:Q1DPXP}(b)]. QMBSs on non-bipartite lattices, such as the triangular lattice, have remained largely unexplored and no scar states have previously been reported in such settings, including the experimental study in Ref.~\cite{bluvstein2021controlling}. Below we demonstrate that the \textsc{ScarFinder} successfully identifies an exact scar eigenstate for the geometry in Fig.~\ref{fig:Q1DPXP}(b).

Specifically, we perform random sampling over initial states and apply \textsc{ScarFinder} using a projection time step of $\Delta t = 1.0$ for 100 iterations. For bond dimensions $\chi > 2$, the algorithm consistently converges to a unique low-entanglement initial state. Upon inspection, we find this state can be exactly represented with $\chi = 2$. Numerical checks confirm that this identified state remains invariant under the PXP Hamiltonian dynamics, revealing itself as an exact eigenstate.

To analyze the structure of the found MPS state, we first note that MPS tensors have nonzero entries only when the physical indices correspond to the three states in Eq.~(\ref{eq:4site}). Thus, the local Hilbert space is effectively reduced to a spin-1 degree of freedom per site. To further characterize the state, we restrict the system to a finite periodic chain of length $L = 4$, convert the MPS into a state vector, and inspect the nonzero amplitudes. We find that these nonzero coefficients all have magnitude $1/\sqrt{17}$ and differ only by sign. The nonzero amplitudes correspond exactly to product-state configurations that avoid placing two adjacent $|\mathbb{Z}_2\rangle$ or $|\mathbb{Z}_2’\rangle$ blocks. For a length-four periodic system, precisely 17 such allowed configurations exist. The sign of each amplitude depends on the parity (even or odd positions) of the $|\mathbb{Z}_2\rangle$ and $|\mathbb{Z}_2’\rangle$ states. Based on these insights, we determine the exact wave function
\begin{equation}
|S\rangle = \mathcal{P}\bigotimes_j \left( |0\rangle_j - |\mathbb{Z}_2\rangle_j - |\mathbb{Z}_2’\rangle_j \right),
\end{equation}
where the projector $\mathcal{P}$ eliminates any configurations containing adjacent $|\mathbb{Z}_2\rangle$ or $|\mathbb{Z}_2’\rangle$ blocks.

Given this form, one can straightforwardly verify that the state is annihilated by the PXP Hamiltonian. In particular, each local term $\hat{h}_j = \mathcal{P}\hat{X}_j\mathcal{P}$, where $\hat{X}_j = \hat{\sigma}_{j,A}^x + \hat{\sigma}_{j,B}^x + \hat{\sigma}_{j,C}^x + \hat{\sigma}_{j,D}^x$ is the sum of Pauli-$X$ operators acting on all sublattice spins at site $j$, individually annihilates the state $|S\rangle$, as shown below:
\begin{equation}
\begin{aligned}
\hat{h}_j |S\rangle 
=&\ \mathcal{P}\!\left[\cdots \otimes \hat{h}_j |0\rangle_{j-1} 
\otimes (|0\rangle_j - |\mathbb{Z}_j\rangle \right. \\
&\ \left. - |\mathbb{Z}'_j\rangle) 
\otimes |0\rangle_{j+1} \otimes \cdots \right] \\
=&\ \mathcal{P}\!\left[\cdots \otimes |0\rangle_{j-1} 
\otimes \hat{X}_j (|0\rangle_j - |\mathbb{Z}_j\rangle \right. \\
&\ \left.- |\mathbb{Z}'_j\rangle) 
\otimes |0\rangle_{j+1} \otimes \cdots \right].
\end{aligned}
\end{equation}
The action of $\hat{X}$ on site $j$ gives
\begin{equation}
\begin{aligned}
&\hat{X} (|0\rangle - |\mathbb{Z}\rangle - |\mathbb{Z}'\rangle) \\
&= -\big(|\downarrow\uparrow\uparrow\uparrow\rangle
+ |\uparrow\downarrow\uparrow\uparrow\rangle
+ |\uparrow\uparrow\downarrow\uparrow\rangle
+ |\uparrow\uparrow\uparrow\downarrow\rangle\big).
\end{aligned}
\end{equation}
Each of these configurations contains adjacent excitations and is therefore projected out by $\mathcal{P}$, which enforces the Rydberg blockade constraint. Consequently, $\hat{h}_j|S\rangle = 0$ for every $j$, implying that the full PXP Hamiltonian annihilates the state:
\begin{equation}
\hat{H}_{\mathrm{PXP}} |S\rangle = 0.
\end{equation}
Thus, $|S\rangle$ is an exact eigenstate with eigenvalue $E = 0$.

We emphasize that the exact cancellations discussed above occur as a special case arising from the $4\times \infty$ cylindrical geometry. It would be interesting to understand if some generalization of this construction may apply to quasi-1D geometries with a larger number of horizontal layers or even in the isotropic 2D limit. Furthermore, we note that our search on a $4\times\infty$ cylinder did not reveal any long-lived dynamical scar states (beyond the exact eigenstate mentioned above), suggesting that robust scar dynamics may not be present in more general non-bipartite systems such as the 2D triangular lattice.

\section{Mixed field Ising model}\label{app:mfi}

\begin{figure}[tb]
    \centering
    \includegraphics[width=\linewidth]{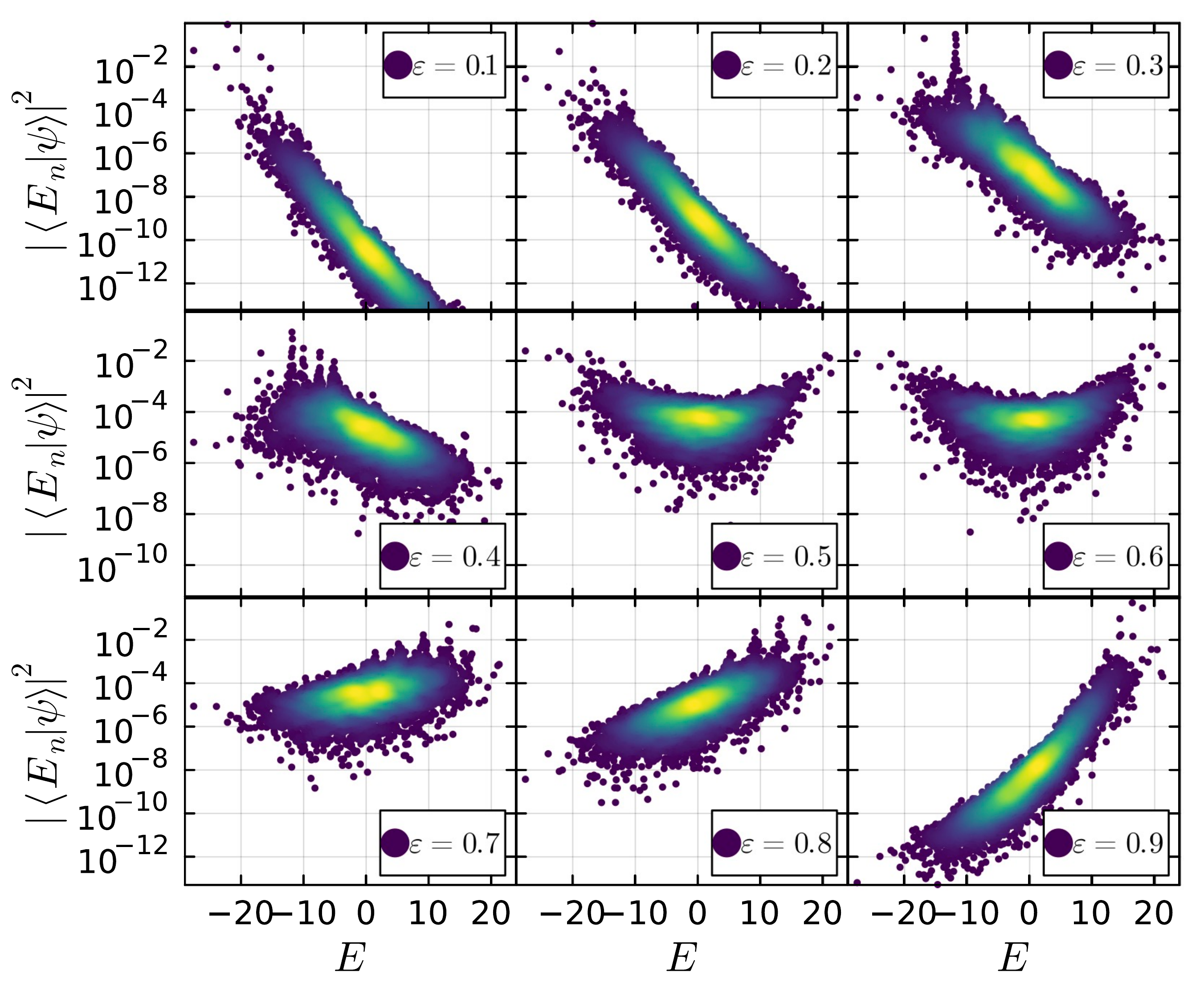}
    \caption{Eigenstate overlaps for optimized initial states obtained via \textsc{ScarFinder} at various target energy densities \(\epsilon\) in the mixed-field Ising model, Eq.~(\ref{eq:isingham}). Each initial state is optimized within the iMPS manifold of bond dimension \(\chi=10\). The optimal states are selected from an ensemble of 100 random initializations, with a time step \(\Delta t=0.5\) over 1000 iterations. The optimal state at each energy density is chosen as the one exhibiting minimal entanglement entropy at \(t=4.0\). The overlap data are obtained via exact diagonalization of the Hamiltonian Eq.~\ref{eq:isingham} (with $(J=1.0, h=0.5, g=1.05$) on a system of size \( L = 16 \), restricted to the momentum sector \( k = 0 \), under two-site translational symmetry. This symmetry reduction arises from the two-site unit cell used in the iTEBD simulation, which explicitly breaks the original one-site translation symmetry down to two-site periodicity.}
    \label{fig:ising_ov}
\end{figure}

The transverse-field Ising model with an additional longitudinal field is a canonical example of a nonintegrable quantum many-body system:
\begin{equation}\label{eq:isingham}
    \hat H_\text{Ising} = -J\sum_j \hat\sigma^z_j \hat\sigma^z_{j+1} - h \sum_j \hat\sigma^z_j - g\sum_j \hat\sigma^x_j.
\end{equation}
For such Hamiltonians, a generic initial state is expected to rapidly thermalize: local observables should swiftly equilibrate to their thermal expectation values, and the bipartite entanglement entropy should quickly saturate to a volume-law scaling.
However, previous studies~\cite{Banuls2011,Alishahiha2025} revealed exceptions to this expectation, where in the chaotic regime (parameters set as \(J=1.0\), \(h=0.5\), \(g=1.05\)), certain initial product states, such as the fully polarized state \(|Z+\rangle \equiv  \left|\uparrow\uparrow\cdots\uparrow\right\rangle\), exhibit anomalously slow thermalization dynamics. This phenomenon, referred to as ``weak thermalization", was subsequently attributed to high overlaps of the initial state with low-entanglement eigenstates in the vicinity of the system's ground state~\cite{LinMotrunich2017}. Recent works, however, have raised questions about whether the model thermalizes in the thermodynamic limit~\cite{PhysRevB.106.214311}, or whether thermal behavior only emerges at much larger system sizes than previously explored~\cite{arxiv:2409.18863}.

Motivated by these observations, we apply the \textsc{ScarFinder} algorithm to systematically search for optimized scar-like initial states in this nonintegrable model. In contrast to our previous investigations of the PXP model in the main text, where it was sufficient to set the targer energy to zero and study the middle of the spectrum, for the mixed-field Ising model in Eq.~(\ref{eq:isingham}) we expect the results to be more sensitive to the energy density. Hence, we vary the target energy density
 $\epsilon = (E - E_0)/(E_\text{max} - E_0)$,
with \(E_0\) and \(E_\text{max}\) denoting the lowest and highest eigenenergies, respectively. Fig.~\ref{fig:ising_ov} shows the eigenstate overlaps of optimized initial states identified via \textsc{ScarFinder} at different energy densities, demonstrating distinctive spectral signatures, especially at higher energy densities.

\begin{figure}[tb]
    \centering
    \includegraphics[width=\linewidth]{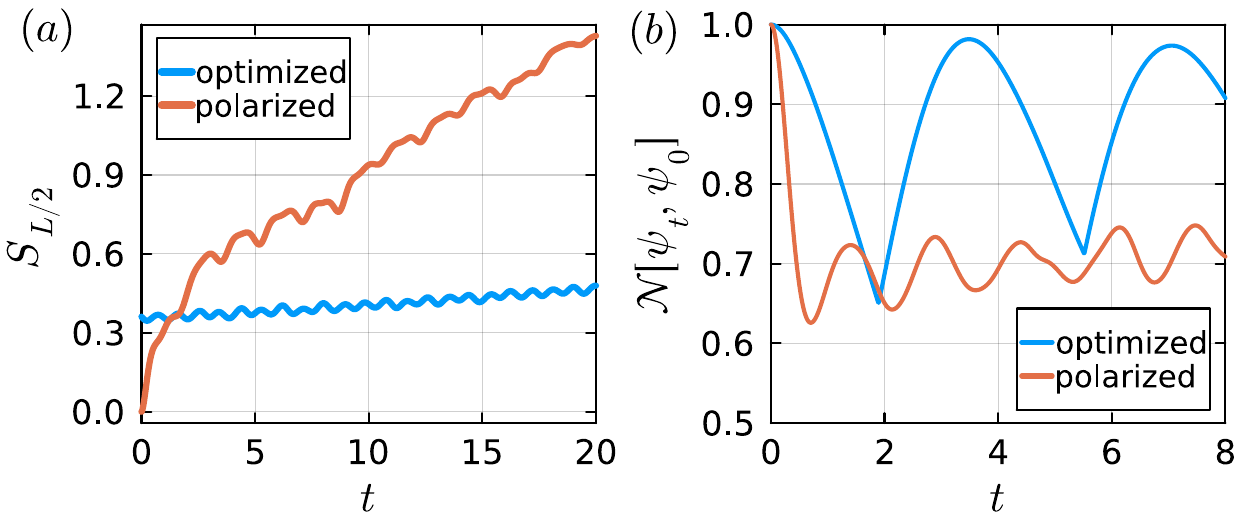}
    \caption{Comparison of the optimized \textsc{ScarFinder} initial state and the polarized product state \( |\theta \approx 0.1649\pi\rangle \) from Eq.~(\ref{eq:productstates}), both at energy density \( \epsilon = 0.1 \). (a) Entanglement entropy growth comparison between the optimized \textsc{ScarFinder} initial state and the product state in Eq.~(\ref{eq:productstates}). The optimized initial state demonstrates significantly suppressed entanglement growth compared to the polarized state. (b) Logarithmic fidelity \( \mathcal{N}[\psi_t, \psi_0] \) as a function of time. The optimized state shows larger and more regular revivals, further confirming its enhanced coherence.}
    \label{fig:eps_0.1}
\end{figure}

To illustrate the improvement provided by \textsc{ScarFinder}, we compare the dynamics of optimized states against simple product states from Ref.~\cite{Banuls2011}, when the two have the same energy density. Specifically, we choose the product state:
\begin{equation}\label{eq:productstates}
    |\theta\rangle = \bigotimes_{j} (\cos\theta\,|{\uparrow_j}\rangle +\sin\theta\,|{\downarrow_j}\rangle),
\end{equation}
with \(\theta \approx 0.1649\pi\), yielding an energy density \(\epsilon=0.1\). Fig.~\ref{fig:eps_0.1}(a) compares the entanglement entropy dynamics from this product state against \textsc{ScarFinder}'s optimized initial state. The optimized state exhibits considerably slower entanglement growth, demonstrating the algorithm's efficacy in isolating dynamically atypical initial conditions with suppressed thermalization. In addition, Fig.~\ref{fig:eps_0.1}(b) shows the logarithmic fidelity dynamics, where the optimized state maintains larger and more regular revivals compared to the polarized state, further supporting its nonthermal character.

The results at intermediate energy density, \(\epsilon \approx 0.5\) in Fig.~\ref{fig:ising_ov}, display characteristics typical of a null result. In particular, the eigenstate overlap distribution shows pronounced weight at both the low- and high-energy ends of the spectrum, forming a superposition reminiscent of an energy-space ``cat state''. Importantly, this does not indicate instability or non-injectivity -- the resulting MPSs are all injective. However, the nonlocal energy distribution suggests that such a state is unlikely to arise from any physically meaningful or experimentally relevant context. This null result, of course, is fully consistent with the expected absence of atypical eigenstates in the middle of the spectrum of a generic chaotic model.

While this demonstrates that the \textsc{ScarFinder} algorithm can successfully identify nonergodic dynamics in generic nonintegrable models, further work is needed to fully understand the underlying mechanisms of these emergent trajectories. In particular, in the intermediate energy density range, e.g., $\epsilon=0.4$ or $\epsilon=0.8$ in Fig.~\ref{fig:ising_ov}, one notices intriguing tower structures that are somewhat reminiscent of the energy-dependent QMBSs in the PXP model in the presence of a chemical potential~\cite{Su2023,Daniel2023}. It would be interesting to understand if these are indeed related to QMBSs or if the quasiparticle picture from Ref.~\cite{LinMotrunich2017} can be generalized to explain these atypical spectral features.

\end{appendix}

\bibliography{ref.bib}

\end{document}